\documentclass[seceq]{ptptex}

\usepackage{graphicx}

\makeatletter
\@ifundefined{lesssim}{\def\lesssim{\mathrel{\mathpalette\vereq<}}}{}
\@ifundefined{gtrsim}{\def\gtrsim{\mathrel{\mathpalette\vereq>}}}{}
\def\vereq#1#2{\lower3pt\vbox{\baselineskip1.5pt \lineskip1.5pt
\ialign{$\m@th#1\hfill##\hfil$\crcr#2\crcr\sim\crcr}}}
\makeatother
\newcommand{\beq}{\begin{equation}}
\newcommand{\eeq}{\end{equation}}
\newcommand{\beqn}{\begin{eqnarray}}
\newcommand{\eeqn}{\end{eqnarray}}

\title{
Stellar core collapse in full general relativity with 
microphysics  \\
-- Formulation and Spheical collapse test --
}

\author{%
Yuichiro \textsc{Sekiguchi} %
}
\inst{%
Division of Theoretical Astronomy,
National Astronomical Observatory of Japan, Mitaka, Tokyo 181-8588,
Japan 
}

\def\bL{\hbox{${\cal L}\llap{ --\,}$}}

\abst{ 

 One of the longstanding issues in numerical relativity is
 to enable a simulation taking account of
 microphysical processes (e.g., weak interactions and neutrino cooling).
 We develop an approximate and explicit scheme in the fully 
 general relativistic framework as a first implementation
 of the microphysics toward a more realistic and sophisticated modeling.
 In this paper, we describe in detail a method for implementation of
 a realistic equation of state, the electron capture and the
 neutrino cooling in a multidimensional, fully general relativistic code.
 The procedure is based on the so-called neutrino leakage scheme.
 To check the validity of the code, we perform a two dimensional (2D)
 simulation of spherical stellar core collapse. 
 Until the convective activities set in, our
 results approximately agree, or at least are consistent, with those
 in the previous so-called state-of-the-art simulations.
 In particular, the radial profiles of thermodynamical quantities and
 the time evolution of the neutrino luminosities agree quantitatively.
 The convection is driven by negative gradients of the entropy per
 baryon and the electron fraction as in the
 previous 2D Newtonian simulations. We clarify which gradient
 is more responsible for the convection. Gravitational waves
 from the convection are also calculated. We find that the
 characteristic frequencies of the gravitational-wave spectra are
 distributed for higher frequencies than those in Newtonian simulations
 due to the general relativistic effects.
}

\begin{document}
\maketitle

%
\section{Introduction}

\subsection{Motivation}

Gravitational collapse of massive stellar core to a neutron 
star or a black hole and the associated supernova explosion are one of the
important and interesting events in the universe.
From observational view point, they are among the most 
energetic events in astrophysics, producing a wide variety of observable
signatures, namely, electromagnetic radiation, neutrinos, and
gravitational radiation.  

Most of the energy liberated in the collapse is eventually carried away
by neutrinos from the system. 
The total energy of neutrinos emitted is 
$\approx GM_{\rm NS}^{2}/R_{\rm NS} \sim 0.1 M_{\rm NS}c^{2}$ $\sim$ several
times $10^{53}$ ergs, where $M_{\rm NS}$ and $R_{\rm NS}$ are the mass
and radius of the neutron star.
Observations of gravitational collapse by neutrino detectors
will provide important information of the deep inside of the core, 
because neutrinos can propagate from the central regions of the stellar
core almost freely due to their small cross-sections with matters. 
Electromagnetic radiation, by contrast,
interacts strongly with matters and thus gives information of collapse
coming only from lower-density regions near the surface of the star. 
Bursts of neutrinos were first detected simultaneously by
the Kamiokande\cite{Hirata87} and Irvine-Michigan-Brookhaven\cite{Bionta87} 
facilities in the supernova SN1987A, 
which occurred on February 23, 1987 in the Large Magellanic Cloud 
(for a review, see Ref.~\citen{Arnett89}).  
Future detection of neutrinos will provide a direct clue to reveal the
physical ingredient for the supernova explosion mechanism. 

Gravitational wave astronomy will start in this decade.
The first generation of ground-based interferometric detectors 
(LIGO\cite{LIGO}, VIRGO\cite{VIRGO}, GEO600\cite{GEO600}, 
TAMA300\cite{TAMA300}) 
are now in the scientific search for gravitational waves. 
Stellar core collapse is one
of the important sources for these observatories.
Observations of gravitational collapse by gravitational-wave detectors
will provide unique information, complementary to that derived from
electromagnetic and neutrino detectors, because gravitational waves can
propagate from the innermost regions of a progenitor star to the detectors
without attenuation by matters.
Combination of the signatures of neutrinos and 
gravitational waves will provide much information about processes of
the core collapse and ultimately, the physics that governs the stellar
core collapse.

To obtain physically valuable information from these observations,
it is necessary to connect the observed data and the physics behind it.
For this purpose, a numerical simulation is the unique approach.
However, simulating the stellar core collapse is one of the challenging 
problems because a rich diversity of physics has to be taken into
account. 
All four known forces in nature are involved and play important
roles during the collapse.
General relativistic gravity plays an essential role in formation of a
black hole and a neutron star.
The weak interactions govern energy and lepton-number losses of the
system. In particular, neutrinos transport most of the energy released
during the collapse to the outside of the system. 
The electromagnetic and strong interactions determine the
ingredient of the collapsing matter and the
thermodynamical properties of the dense matter.
Strong magnetic fields, if they are present, 
would modify the dynamics of the collapse,  subsequent supernova
explosion, and evolution of proto-neutron stars.

Due to these complexities, the explosion mechanism of core-collapse 
supernovae has not been fully understood in spite of the elaborate
effort in the past about 40 years\cite{Kotake06,Janka07a,Bethe90}.
Recent numerical studies\cite{Rampp00,Mezza01,Thompson03,RJ02,Lieben01,Sumi05} 
have clarified that on the assumption of the spherical symmetry, 
the explosion does not succeed for the iron core collapse 
with the currently most elaborate input physics 
(neutrino interactions, neutrino transfer, and equation of states of the dense matter) 
on the basis of the standard ``neutrino heating mechanism''\cite{Bethe90}
(but see Ref.~\citen{Kitaura06} for successful explosion 
in O-Ne-Mg core collapse).
To increase the neutrino-heating efficiency, a wide variety of
multi-dimensional effects have been explored (for recent reviews, see
e.g., Refs.~\citen{Janka07a}, \citen{Kotake06} and also
Refs.~\citen{MJ09} and \citen{Burrows06a} for simulations where
successful explosions are obtained). However, it has not been completely
clarified yet whether the increase of the heating efficiency due to such
multi-dimensional effects suffices for yielding successful explosion,
because the explosion energy resulting from these works is too low 
$\sim 10^{50}$ ergs.

Similarly, accurate predictions of gravitational waveforms are still 
hampered by the facts that reliable estimates of waveforms require a general 
relativistic treatment\cite{Dimm02}, and that appropriate treatments of 
microphysics such as nuclear equation of state (EOS), the electron capture,
and neutrino emissions and transfers. 
Indeed, previous estimates of waveforms have 
relied either on Newtonian simulations with including microphysics to some 
extent\cite{Monch91,Burrows96,MJ97,Zwerg97,Kotake,Ott04,Muller04,Fryer04}, 
or general relativistic simulations with simplified 
microphysics\cite{Dimm02,SS,Sekiguchi05,Duran05,Shibata06}.
Recently, gravitational waveforms emitted in the rotating core collapse
were derived by multidimensional simulations in general relativistic
frameworks\cite{Ott07,Dimm07} adopting a finite-temperature nuclear 
EOS\cite{Shen98} and the electron capture. 
In their studies, however, the electron capture rate was not
calculated in a self-consistent manner. Instead, they adopted
a simplified prescription proposed in Ref.~\citen{Lieb05} which is 
based on the result of a spherically symmetric simulation.
However, it is not clear whether this treatment is justified for
non-spherical collapse or not.
Moreover, they did not take account of emission processes of neutrinos.
More sophisticated simulations including microphysics are required to make
accurate predictions of gravitational waveforms.

The gravitational collapse of massive star is also
the primary mechanism of black hole formation.
Understanding the process of black hole formation is one of 
the most important issues in the theory of the stellar core collapse. 
A wide variety of recent observations
have shown that black holes actually exist in the universe
(e.g., see Ref.~\citen{Rees03}), and so far, 
about 20 stellar-mass black holes for which
the mass is determined within a fairly small error have been
observed in binary systems of our Galaxy and the Large Magellanic
Clouds\cite{McC06}. 
The formation of a black hole through the gravitational collapse 
is a highly nonlinear and dynamical phenomenon, and therefore,
numerical simulation in full general relativity is the unique 
approach to this problem.
In spherical symmetry, fully general relativistic simulations of stellar
core collapse to a black hole have been performed in
a state-of-the-art manner, i.e., employing realistic EOSs, 
implementing microphysical processes, and the Boltzmann transfer 
of neutrinos\cite{Sumi06,Nakazato}.
In the multidimensional case, by contrast, simulations only with
simplified microphysics have been performed
\cite{Sekiguchi05,Sekiguchi07,Liu}.
Because multidimensional effects such as rotation and convection are
likely to play an important role, multidimensional simulations in full 
general relativity employing a realistic EOS and detailed microphysics 
are necessary for clarifying the formation process of black holes.

Furthermore, recent observations\cite{980425,030329,060218} 
have found the spectroscopic connections between 
several SNe and long gamma-ray bursts (GRBs) and clarified that some of long
GRBs are associated with the collapse of massive stars.
Supported by these observations, the collapsar model\cite{Collapsar} 
is currently one of promising models for the central engine of GRBs. 
In this model, one assumes that the central engine of the long GRBs 
is composed of a rotating black hole and a hot, massive accretion disk. 
Such a system may be formed as a result of the collapse of
rapidly rotating massive core.
In this model, one of the promising processes of the energy-deposition
to form a GRB fireball is the pair 
annihilation of neutrinos emitted from the hot, massive disk 
($ \nu_{e} + \bar{\nu}_{e} \rightarrow e^{-} + e^{+} $).
The collapsar model requires the progenitor core to be rotating rapidly
enough that the massive accretion disk can be formed around the black hole.
Recent general relativistic numerical analyses have shown that if a 
progenitor of the collapse is massive and the angular momentum is large
enough, a black hole surrounded by a massive disk will be 
formed\cite{Shibata02,Sekiguchi04,Sekiguchi07}.
However, the formation mechanism of such system has not
been clarified in detail.
These also enhance the importance of exploring the stellar core collapse 
to a black hole taking account of microphysical processes. 

As reviewed above, multidimensional simulations of stellar collapse 
in full general relativity including microphysics is currently one of 
the most required subjects in theoretical astrophysics. 
However, there has been no multidimensional code in full general relativity 
that self-consistently includes microphysics such as
realistic EOS, electron capture, and neutrino emission.
There have only existed fully general relativistic codes in spherical
symmetry\cite{Yamada99,Lieb04,Sumi05} or Newtonian codes in
multidimension\cite{Rampp00,Mezza01,Thompson03}.
We have developed a fully general relativistic multidimensional 
code including a finite-temperature nuclear EOS, 
self-consistent treatment of the electron capture, 
and a simplified treatment of neutrino emission for the first time.
In this code, by contrast with the previous ones\cite{Ott07,Dimm07}, 
the electron capture rate is treated in a self-consistent manner and the
neutrino cooling is taken into account for the first time.
Because it is not currently feasible to fully solve the 
neutrino transfer equations in the framework of general relativity 
in multidimension because of restrictions of computational resources,
it will be reasonable to take some approximation for the transfer
equations at the current status.
In this paper, the so-called neutrino leakage scheme is adopted as 
an approximate treatment of neutrino cooling, and a general
relativistic version of the leakage scheme is developed.

\subsection{The leakage schemes}

The neutrino leakage
schemes\cite{EP81,vRL81,vanRiper82,BLH82,RCK84,BCK85,Cooperstein88,Kotake03} 
as an approximate method for the neutrino cooling have a well-established 
history (e.g.\ Ref.~\citen{Cooperstein88}). The basic concept of the 
original neutrino leakage schemes\cite{EP81,vRL81} is to treat the 
following two regions in the system separately: one is the region where 
the diffusion timescale of neutrinos is longer than the dynamical 
timescale, and hence, neutrinos are 'trapped' (neutrino-trapped region); 
the other is the region where the diffusion timescale is shorter than the 
dynamical timescale, and hence, neutrinos stream out freely out of the 
system (free-streaming region). 
The idea of treating the diffusion region separately has been applied to 
more advanced methods for the neutrino transfer
(see e.g., Ref.~\citen{Ott08} and references therein).

Then, electron neutrinos and electron anti-neutrinos in the
neutrino-trapped region are assumed to be in the $\beta$-equilibrium state. 
The {\it net} local rates of lepton-number and energy exchange with matters 
are set to be zero in the neutrino-trapped region.
To treat diffusive emission of neutrinos leaking out of the 
neutrino-trapped region, simple phenomenological source terms based on 
the diffusion theory are introduced\cite{EP81,vRL81}.
In the free-streaming region, on the other hand, it is assumed that 
neutrinos escape from the system without interacting with matters. 
Therefore, neutrinos carry the lepton number and the energy according to
the local weak-interaction rates. 
Note that the neutrino fractions are not solved in the original 
version of the leakage scheme: Only the total lepton fraction (from
which the neutrino fractions are calculated under the
$\beta$-equilibrium condition) is necessary in the neutrino-trapped
region, and the neutrino fractions are set to be zero in the free-streaming 
region. As a result, neutrino quantities and the electron fraction are 
discontinuous at the boundary the neutrino-trapped and free-streaming
regions. 

The boundary was given by hand as a single 'neutrino-trapping' 
density ($\rho_{\rm trap}$) without calculating the optical depths of
neutrinos in the previous studies
\cite{EP81,vRL81,vanRiper82,BLH82,RCK84,Kotake03}. 
However, the location at which the neutrino trapping occurs in fact 
depends strongly on the neutrino energies ($E_{\nu}$)
as\cite{Bethe1990} 
$\rho_{\rm trap} \propto E_{\nu}^{\ -3}$, and hence, there are
different neutrino-trapping densities for different neutrino energies. 
In the previous leakage schemes
\cite{EP81,vRL81,vanRiper82,BLH82,Kotake03}, 
on the other hand, all neutrinos were emitted in one moment 
irrespective of their energy.
Consequently in the case of the so-called neutrino burst
emission (e.g., Ref.~\citen{Bethe1990}), for example, 
the duration in which the neutrinos are emitted was shortened and 
the peak luminosity at the burst was overestimated\cite{vanRiper82,Kotake03,PhD}.
The dependence of the neutrino-trapping densities and the neutrino 
diffusion rates on the neutrino energies are approximately taken 
into account in the recent simulations of mergers of binary neutron 
star\cite{RJS96,RL03}. 
However, the lepton-number conservation equations for neutrinos are 
not solved\cite{RJS96}, which is important to estimate the phase space
blocking due to neutrinos. 

{\it Transfer equations} of neutrinos are not solved in the leakage
schemes. Therefore, the leakage schemes cannot treat {\it non-local} 
interactions among the neutrinos and matters. For example, 
the so-called neutrino heating\cite{BW85} and the neutrino pair 
annihilation cannot be treated in the leakage scheme.
Nevertheless, we believe a detailed general relativistic leakage
scheme presented in this paper to be a valuable approach 
because even by this approximated approach it is possible to incorporate
the effects of neutrinos semi-quantitatively as shown in this paper.
Also, the neutrino leakage scheme is an appropriate method for
studying a number of phenomena for which the neutrino heating and
neutrino transfer are expected to be not very important, e.g., 
prompt formation of a black hole and compact binary mergers.
Both of these phenomena are the targets of the present code.

A first attempt towards a general relativistic leakage scheme was 
done in the previous study\cite{PhD}.
In that study, not the region of the system but the energy momentum 
tensor of neutrinos was decomposed into two parts;
'trapped-neutrino' and 'streaming-neutrino' parts. 
However the source terms of hydrodynamic and 
lepton-number-conservation equations were determined using the single 
neutrino-trapping density as in the case of the previous leakage 
schemes. 
In this paper, we develop a new code implementing the microphysical
processes in the general relativistic framework based on the previous
study\cite{PhD}.
As an application of the code, we perform simulations of stellar
core collapse.

A lot of improved ingredients are installed into the present code:
(1) The dependence of the neutrino diffusion rates on the neutrino 
energies are approximately taken into account following the recent study
\cite{RL03} with detailed cross sections, 
instead of adopting the single neutrino-trapping density (see Appendix C).
(2) The lepton-number conservation equations for neutrinos are solved\
to calculate self-consistently the chemical potentials of neutrinos.
Then, the blocking effects due to the presence of neutrinos and the
$\beta$-equilibrium condition can be taken into account more accurately
(see \S \ref{GRleak}).
(3) A stable explicit method for solving the equations of hydrodynamics,
the lepton-number conservations, and neutrinos are developed. 
Such a special procedure is necessary because the characteristic 
timescale of the weak-interaction processes 
(hereafter referred to as the WP timescale 
$t_{\rm wp} \sim \vert Y_{e}/\dot{Y}_{e} \vert $) 
is much shorter than the dynamical timescale $t_{\rm dyn}$ in hot,
dense matter regions\cite{Bruenn85,RL03}.
Note that in the previous leakage
schemes\cite{EP81,vRL81,vanRiper82,BLH82,Kotake03}, the
$\beta$-equilibrium was assumed to be achieved in such regions 
(i.e. $\dot{Y}_{e} = 0$) and no such special treatments are required.
See \S \ref{Difficulty} for further discussions and \S \ref{GRleak}
for details of the method.
(4) The electron capture rate are calculated in a detailed
manner\cite{FFN85} including effects of the so-called thermal 
unblocking\cite{CW84} (see Appendix A).

The paper is organized as follows. 
First, issues in implementation of weak interactions and 
neutrino cooling in full general relativistic simulation 
is briefly summarized in \S \ref{Difficulty}.
Then, framework of the general relativistic leakage scheme is described 
in detail in \S~\ref{GRleak}.
In \S~\ref{S_EOS}, EOSs employed in this paper are described in some details.
Basic equations and numerical methods of the simulations are 
described in \S~\ref{S_Numerical}. 
Numerical results obtained in this paper are shown in \S~\ref{S_Results}.
We devote \S~\ref{S_Summary} to a summary and discussions.
In appendices, details of the microphysics adopted in the present paper
are summarized for the purpose of convenience.
Throughout the paper, the geometrical unit $c=G=1$ is used otherwise
stated.

\section{Issues in implementation of weak interactions and
 neutrino cooling in fully general relativistic simulation}\label{Difficulty}

Because the characteristic timescale of the weak-interaction processes 
(the WP timescale $t_{\rm wp} \sim \vert Y_{e}/\dot{Y}_{e} \vert $) 
is much shorter than the dynamical timescale $t_{\rm dyn}$ in hot 
dense matters\cite{Bruenn85,RL03}, the {\it explicit} numerical
treatments of the weak interactions are computationally expensive 
in simple methods, as noted in the previous pioneering work 
by Bruenn\cite{Bruenn85}:
A very short timestep ($\Delta t$ $<$ $t_{\rm wp} \ll t_{\rm dyn}$) will
be required to solve the equations explicitly.

The {\it net} rates of lepton-number and energy exchanges between 
matters and neutrinos may not be large, and consequently, an 
{\it effective} timescale may not be as short as the dynamical
timescale. However, this does not immediately imply that one can solve the
equations explicitly without employing any prescription.
For example, the achievement of $\beta$-equilibrium, where $\dot{Y}_{e}=0$ 
is the consequence of the cancellation of two very {\it large} weak 
interaction processes (the electron and the electron-neutrino captures,
see Eq. (\ref{sourceYe})) 
and of the action of the phase space blocking.
Note that the weak interaction processes depend enormously both on the 
temperature and the lepton chemical potentials.
Therefore, small error in the evaluation of the temperature and 
a small deviation from the $\beta$-equilibrium due to small error in
calculation of the lepton chemical potentials will result in huge 
error. Then, stiff source terms appear and explicit numerical
evolution often becomes unstable. 
Indeed, we found that a straightforward, explicit solution of the 
equations did not work.

In the following of this section, we describe issues of
implementation of weak interactions and neutrino cooling into the 
hydrodynamic equations in the conservative schemes in fully general
relativistic simulations.
Fiest, we illustrate that
in the Newtonian framework, the equations may be solved implicitly
in a relatively simple 
manner\cite{BW82,Bruenn85,MBHLSR87,MB93,RJ02,Livne04,Buras06,Burrows07,MJ09}
(see also Refs.~\citen{MM99} and \citen{Ott08} and references therein). 
The equations of hydrodynamics, lepton-number conservations, and neutrino
processes are schematically written as,
\beqn
\dot{\rho   } &=& 0 , \\
\dot{v_{i}  } &=& S_{v_{i}  }(\rho, Y_{e}, T, Q_{\nu}) , \\
\dot{Y_{e}  } &=& S_{Y_{e}  }(\rho, Y_{e}, T, Q_{\nu}) , \\
\dot{e      } &=& S_{e      }(\rho, Y_{e}, T, Q_{\nu}) , \\
\dot{Q_{\nu}} &=& S_{Q_{\nu}}(\rho, Y_{e}, T, Q_{\nu}) , 
\eeqn
where $\rho$ is the rest-mass density, $v_{i}$ is the velocity, 
$Y_{e}$ is the electron fraction,
$e$ is the (internal) energy of matter, $T$ is the temperature, and 
$Q_{\nu}$ stands for the relevant neutrino quantities. 
We here omit the transport terms.
$S$'s in the right-hand side stand for the relevant source terms. 
Comparing the quantities in the left-hand-side and the argument quantities 
in the source terms, only the relation between $e$ and $T$ is nontrivial.
Usually, EOSs employed in the simulation are tabularized, and 
one dimensional search over the EOS table is required to solve them.
Due to the relatively simple relations between the quantities to be 
evolved and the argument quantities, the above equations may be solved 
implicitly in a straightforward (although complicated) manner.

In the relativistic framework, the situation becomes much more
complicated in conservative schemes, because the Lorentz factor ($\Gamma$)
is coupled with rest-mass density and the energy density (see
Eqs. (\ref{continuS}) and (\ref{eneS}) where 
$w \equiv \alpha u^{t}$ is used instead of $\Gamma$), 
and because the specific enthalpy
($h = h(\rho,Y_{e},T)$) is coupled with the momentum (see
Eq. (\ref{momS})).

It should be addressed that the previous fully general relativistic
works in the spherical symmetry\cite{Yamada99,Lieb04} are based on 
the so-called Misner-Sharp coordinates\cite{MS64}.
There are no such complicated couplings in these {\it Lagrangian} 
coordinates.
Accordingly, the equations may be solved essentially in the same
manner as in the Newtonian framework. 
Because no such simple Lagrangian coordinates are known in the
multidimensional case, the complicated couplings inevitably appear 
in the relativistic framework.

Omitting the factors associated with the geometric variables 
(which are usually updated before solving the hydrodynamics equations)
and the transport terms, 
the equations to be solved in the general relativistic framework 
are schematically written as,
\beqn
\dot{\rho_{*}}(\rho,\Gamma)  &=& 0 , \label{rhoEq} \\
\dot{\hat{u}}_{i}(u_{i},h) = \dot{\hat{u}}_{i}(u_{i},\rho,Y_{e},T) 
        &=& S_{\hat{u}_{i}}(\rho, Y_{e}, T, Q_{\nu}, \Gamma) , \\
\dot{Y_{e}  } &=& S_{Y_{e}  }(\rho, Y_{e}, T, Q_{\nu}, \Gamma) , \\
\dot{\hat{e}}(\rho, Y_{e}, T, \Gamma) &=& 
S_{\hat{e}}(\rho, Y_{e}, T, Q_{\nu}, \Gamma) , \\
\dot{Q_{\nu}} &=& S_{Q_{\nu}}(\rho, Y_{e}, T, Q_{\nu}, \Gamma) , 
\eeqn
where $\rho_{*}$ is a weighted density, $\hat{u}_{\alpha}$ is a weighted 
four velocity, $\hat{e}$ is a weighted energy density (see \S~\ref{BasicEq} 
for the definition of these variables).
The Lorentz factor is calculated by solving the normalization condition 
$u^{\alpha}u_{\alpha}=-1$, which is rather complicated nonlinear equation 
schematically written as
\beq
f_{\rm normalization}(\hat{u_{i}}, \Gamma) 
= f_{\rm normalization}(u_{i}, \rho, Y_{e}, T, \Gamma) = 0. \label{nomEq}
\eeq
The accurate calculation of the Lorentz factor and the accurate solution
of the normalization condition are very important in the numerical
relativistic hydrodynamics.

Now, it is obvious that the argument quantities in the source terms are
not simply related with the evolved quantities in the left-hand-side of Eqs. 
(\ref{rhoEq})--(\ref{nomEq}). Solving the equations implicitly is not
as straightforward as in the Newtonian case and no successful
formulations have been developed.
Moreover it might be not clear whether a convergent solution can be 
{\it stably} obtained numerically or not, because they are simultaneous 
nonlinear equations. Therefore, it may be not a poor choice to adopt 
an alternative approach in which the equations are solved 
{\it explicitly} with some approximations as described in the next
section\footnote{It should be stated that
the implicit schemes are also approximated ones because
a short WP timescale associated with the weak interaction 
is not fully resolved.}.

\section{General relativistic neutrino leakage scheme}\label{GRleak}

In this section, we describe a method for approximately 
solving hydrodynamic equations coupled with neutrino radiation 
in an explicit manner.
As described in the previous section, because of the relation
$t_{\rm wp} \ll t_{\rm dyn}$ in the hot dense matter regions, 
the source terms in the equations become too {\it stiff} for the
equations to be solved explicitly in the straightforward manner. 
The characteristic timescale of leakage of neutrinos
from the system $t_{\rm leak}$, by contrast, is much longer 
than $t_{\rm wp}$ in the hot dense matter region. 
Rather, $t_{\rm leak} \sim L/c \sim t_{\rm dyn}$ where $L$ is the
characteristic length scale of the system. On the other hand,
$t_{\rm leak}$ is comparable to $t_{\rm wp}$ in the free-streaming
regions but $t_{\rm wp}$ is longer than or comparable 
with $t_{\rm dyn}$ there. All these facts imply that the WP timescale
does not directly determine the evolution of the system but the leakage
timescale does.
Using this fact, we approximate some of original equations and 
reformulate them so that the source terms are to be characterized by 
the leakage timescale $t_{\rm leak}$.

\subsection{Decomposition of neutrino energy-momentum
  tensor}\label{EnergyMomentum}

The basic equations of general relativistic hydrodynamics with neutrinos
are
\beq
\nabla_{\alpha}(T^{\rm Total})^{\alpha}_{\beta} 
= \nabla_{\alpha}\left[(T^{\rm F})^{\alpha}_{\beta} +
(T^{\nu})^{\alpha}_{\beta} \right] = 0, \label{eqtot}
\eeq
where $(T^{\rm Total})_{\alpha \beta}$ is the total energy-momentum
tensor, and $(T^{\rm F})_{\alpha \beta}$ and $(T^{\nu})_{\alpha \beta}$
are the energy-momentum tensor of fluids and neutrinos, respectively.
Equation (\ref{eqtot}) can be written as
\beqn
\nabla_{\alpha}(T^{{\rm F}})^{\alpha}_{\beta} &=& Q_{\beta}, \label{T_Eq1} \\
\nabla_{\alpha}(T^{\nu})^{\alpha}_{\beta} &=& -Q_{\beta} \label{T_Eq2},
\eeqn
where the source term $Q_{\alpha}$ is regarded as the local production 
rate of neutrinos through the weak processes.

Now, the problem is that the source term $Q_{\alpha}$ becomes 
too stiff to solve explicitly in hot dense matter regions where 
$t_{\rm wp} \ll t_{\rm dyn}$.
To overcome this situation, the following procedures are adopted.

First, it is assumed that the energy-momentum tensor of neutrinos are
be decomposed into 
'trapped-neutrino' ($(T^{\nu,{\rm T}})_{\alpha\beta}$) 
and 'streaming-neutrino' ($(T^{\nu,{\rm S}})_{\alpha\beta}$) parts as
\cite{PhD},
\beq
(T^{\nu})_{\alpha\beta} = (T^{\nu,{\rm T}})_{\alpha\beta} +
                       (T^{\nu,{\rm S}})_{\alpha\beta} . 
\label{nudecompose}
\eeq
Here, the trapped-neutrinos phenomenologically represent neutrinos which 
interact sufficiently frequently with matter and are thermalized.
On the other hand, the streaming-neutrino part describes a
phenomenological flow of 
neutrinos streaming out of the system \cite{PhD} 
(see also Ref.~\citen{Lieb09} where a more sophisticate method in terms 
of the distribution function is adopted in the Newtonian framework).

Second, the locally produced neutrinos are assumed 
to {\it leak out} to be the streaming-neutrinos
with a leakage rate $Q^{\rm leak}_{\alpha}$:
\beq
\nabla_{\beta}(T^{\nu,{\rm S}})^{\beta}_{\alpha} =  Q^{\rm leak}_{\alpha}.
\label{T_Eq_nuS}
\eeq
Then, the equation of the trapped-neutrino part is
\beq
\nabla_{\beta}(T^{\nu,{\rm T}})^{\beta}_{\alpha} = 
Q_{\alpha} - Q^{\rm leak}_{\alpha}.
\label{T_Eq_nuT}
\eeq

Third, the trapped-neutrino part is combined with the fluid part as
\beq
T_{\alpha\beta} \equiv (T^{\rm F})_{\alpha\beta} 
+ (T^{\nu,{\rm T}})_{\alpha\beta},
\eeq
and Eqs. (\ref{T_Eq1}) and (\ref{T_Eq_nuT}) are combined to give
\beq
\nabla_{\beta}T^{\beta}_{\alpha} = -Q^{\rm leak}_{\alpha} \label{T_Eq_M}.
\eeq
Thus, the equations to be solved are changed from 
Eqs. (\ref{T_Eq1}) and (\ref{T_Eq2}) to 
Eqs. (\ref{T_Eq_M}) and (\ref{T_Eq_nuS}).
Note that the new equations only include the source term 
$Q^{\rm leak}_{\alpha}$ which is characterized by the leakage 
timescale $t_{\rm leak}$.
Definition of $Q^{\rm leak}_{\alpha}$ will be given 
in \S~\ref{leakagerate}.

The energy-momentum tensor of the fluid and trapped-neutrino parts
($T_{\alpha \beta}$) is treated as that of the perfect fluid,
\beq
T_{\alpha\beta} = (\rho + \rho \varepsilon + P)
 u_{\alpha}u_{\beta} + P g_{\alpha\beta}, \label{T_fluid}
\eeq
where $\rho$ and $u^{\alpha}$ are the rest mass density and the 4-velocity.
The specific internal energy density ($\varepsilon$) and the pressure ($P$) 
are the sum of contributions from the baryons 
(free protons, free neutrons, $\alpha$-particles, and heavy nuclei), 
leptons (electrons, positrons, and {\it trapped-neutrinos}), and the photons as,
\beqn
P &=& P_{B} + P_{e} + P_{\nu} + P_{ph}, \\
\varepsilon &=&  
\varepsilon_{B} + \varepsilon_{e} +
\varepsilon_{\nu} + \varepsilon_{ph} ,
\eeqn
where subscripts '$B$', '$e$', '$ph$', and '$\nu$' denote the components
of the baryons, electrons and positrons, photons, and trapped-neutrinos, 
respectively. 

The streaming-neutrino part, on the other hand, is set to be a general
form of
\beq
(T^{\nu,{\rm S}})_{\alpha\beta}= 
E n_{\alpha}n_{\beta} + F_{\alpha}n_{\beta} + F_{\beta}n_{\alpha} + P_{\alpha\beta},
\label{T_neutrino}
\eeq
where $F_{\alpha}n^{\alpha}=P_{\alpha \beta}n^{\alpha}=0$. 
In order to close the system,
we need an explicit expression of $P_{\alpha \beta}$.
In this paper, we adopt a simple form 
$P_{\alpha \beta}=\chi E \gamma_{\alpha \beta}$ with $\chi = 1/3$.
This approximation may work well in high density regions but
will violate in low density regions. However, the violation will 
not affect the dynamics because the total amount of streaming-neutrinos 
emitted in low density regions will be small.
Of course, a more sophisticated treatment will be necessary in a future
study.

\subsection{The lepton-number conservation equations}\label{Lepton}

The conservation equations of the lepton fractions are written schematically as
\beqn
&&\!\! \frac{d Y_{e}}{dt} = -\gamma_{e} , \label{dYe} \\
&&\!\! \frac{d Y_{\nu e}}{dt} = \gamma_{\nu e},  \label{dYnu} \\ 
&&\!\! \frac{d Y_{\bar{\nu} e}}{dt} = \gamma_{\bar{\nu} e},  \label{dYna} \\ 
&&\!\! \frac{d Y_{\nu x}}{dt} = \gamma_{\nu x},  \label{dYno}  
\eeqn
where $Y_{e}$, $Y_{\nu e}$, $Y_{\bar{\nu} e}$, and $Y_{\nu x}$ denote
the electron fraction, the electron neutrino fraction, the electron
anti-neutrino fraction, and $\mu$ and $\tau$ neutrino and anti-neutrino 
fractions, respectively. 
We note that in the previous simulations based on the leakage schemes 
\cite{EP81,vRL81,Kotake03,RJS96}, the neutrino fractions 
were not solved.

The source terms of neutrino fractions can be written, on the basis of 
the present leakage scheme, as 
\beqn
&&\!\! \gamma_{\nu e} = \gamma_{\nu e}^{\rm local} - \gamma_{\nu e}^{\rm leak}, \\
&&\!\! \gamma_{\bar{\nu} e} = \gamma_{\bar{\nu} e}^{\rm local} 
                        - \gamma_{\bar{\nu} e}^{\rm leak}, \\
&&\!\! \gamma_{\nu x} = \gamma_{\nu x}^{\rm local} - \gamma_{\nu x}^{\rm leak}, 
\eeqn
where $\gamma^{\rm local}$'s and $\gamma^{\rm leak}$'s are
the local production and the leakage rates of each neutrino, respectively
(see \S~\ref{leakagerate}).
Note that only the trapped-neutrinos are responsible for 
the neutrino fractions.
Assuming that the trapped neutrinos are thermalized and the distribution
function is the equilibrium Fermi-Dirac one, the chemical
potentials of neutrinos can be calculated 
from the neutrino fractions. Then the thermodynamical quantities 
of neutrinos can be also calculated. 

The source term of the electron fraction conservation is
\beq
\gamma_{e} = \gamma_{\nu e}^{\rm local} - \gamma_{\bar{\nu} e}^{\rm local}.\label{sourceYe}
\eeq
Because $\gamma^{\rm local}_{\nu}$\ 's are characterized 
by the WP timescale $t_{\rm wp}$, some procedures are necessary to solve
the lepton conservation equations explicitly.
The following simple procedures are proposed to solve the equations stably. 

First, in each timestep $n$, the conservation equation of 
the {\it total} lepton fraction ($Y_{l}=Y_{e}-Y_{\nu e}+Y_{\bar{\nu} e}$), 
\beqn
&&\!\! \frac{d Y_{l}}{dt} = -\gamma_{l},  \label{dYl} 
\eeqn
is solved together with the conservation equation of $Y_{\nu x}$, Eq. (\ref{dYno}),
in advance of solving whole of the lepton conservation 
equations (Eqs. (\ref{dYe}) -- (\ref{dYno})).
Note that the source term 
$\gamma_{l} = \gamma_{\nu e}^{\rm leak} - \gamma_{\bar{\nu} e}^{\rm leak}$ 
is characterized by the leakage timescale $t_{\rm leak}$
so that this equation can be solved explicitly in the hydrodynamic timescale.
Then, assuming that the $\beta$-equilibrium is achieved, 
values of the lepton fractions in the $\beta$-equilibrium ($Y_{e}^{\beta}$,
$Y_{\nu e}^{\beta}$, and $Y_{\bar{\nu} e}^{\beta}$) are calculated from
the evolved $Y_{l}$. 

Second, regarding $Y_{\nu e}^{\beta}$ and $Y_{\bar{\nu} e}^{\beta}$ as the
maximum allowed values of the neutrino fractions in the next 
timestep $n+1$,  the source terms are limited so that $Y_{\nu}$'s in 
the timestep $n+1$ cannot exceed $Y_{\nu}^{\beta}$ 's.
Then, the whole of the lepton conservation equations 
(Eqs. (\ref{dYe}) -- (\ref{dYno})) are solved explicitly using these limiters.

Third, the following conditions are checked 
\beqn
\mu_{p}+\mu_{e} < \mu_{n}+\mu_{\nu e} , \\
\mu_{n}-\mu_{e} < \mu_{p}+\mu_{\bar{\nu} e},
\eeqn
where $\mu_{p}$, $\mu_{n}$, $\mu_{e}$, $\mu_{\nu e}$ and $\mu_{\bar{\nu}
e}$ are the chemical potentials of protons, neutrons, electrons, 
electron neutrinos, and electron anti-neutrinos, respectively. 
If both conditions are satisfied, the values of
the lepton fractions in the timestep $n+1$ are set to be those in 
the $\beta$-equilibrium value;  
$Y_{e}^{\beta}$, $Y_{\nu e}^{\beta}$, and $Y_{\bar{\nu} e}^{\beta}$.
On the other hand, if either or both conditions are not satisfied, 
the lepton fractions in the timestep $n+1$ is set to be those obtained
by solving whole of the lepton-number conservation equations.  

A limiter for the evolution of $Y_{\nu x}$ may be also necessary 
in the case where the pair processes are dominant, for example, 
in the simulations for collapse of population III stellar core.
In this case, the value of $Y_{\nu x}$ at the pair equilibrium 
(i.e. at $\mu_{\nu x}=0$), $Y_{\nu x}^{\rm pair}$ may be used 
to limit the source term.

\subsection{Definition of leakage rates}\label{leakagerate}

In this subsection the definitions of the leakage rates $Q_{\alpha}^{\rm
leak}$ and $\gamma_{\nu}^{\rm leak}$ are presented.
Because $Q^{\rm leak}_{\nu}$ may be regarded as the emissivity of
neutrinos measured in the {\it fluid rest frame}, $Q^{\rm
leak}_{\alpha}$ is defined as \cite{SSR07}
\beq
Q^{\rm leak}_{\alpha} = Q^{\rm leak}_{\nu}u_{\alpha}.
\eeq\label{leakage_source_Q}

The leakage rates $Q^{\rm leak}_{\nu}$ and 
$\gamma^{\rm leak}_{\nu}$ are assumed to satisfy the following properties.
\begin{enumerate}
\item The leakage rates approach the local rates 
  $Q_{\nu}^{\rm local}$ and $\gamma_{\nu}^{\rm local}$ in the low density, 
  transparent region. 
\item The leakage rates approach the diffusion rates 
  $Q_{\nu}^{\rm diff}$ and $\gamma_{\nu}^{\rm diff}$ in the high density,
  opaque region. 
\item The above two limits should be connected smoothly.
\end{enumerate}
Here, the local rates can be calculated based on the theory of weak 
interactions and the diffusion rates can be determined based on the 
diffusion theory (see appendices for the definition of the local and 
diffusion rates adopted in this paper). 
There will be several prescriptions to satisfy the requirement (iii)
\cite{RJS96,RL03}.
In this paper, the leakage rates are defined by
\beqn
&&\!\! Q_{\nu}^{\rm leak}= (1-e^{-b\tau_{\nu}}) Q_{\nu}^{\rm diff} 
+ e^{-b\tau_{\nu}} Q_{\nu}^{\rm local}, \label{Q_leak} \\
&&\!\! \gamma_{\nu}^{\rm leak}= (1-e^{-b\tau_{\nu}}) \gamma_{\nu}^{\rm diff} 
+ e^{-b\tau_{\nu}} \gamma_{\nu}^{\rm local}, \label{g_leak}
\eeqn
where $\tau_{\nu}$ is the optical depth of neutrinos and $b$ is a parameter
which is typically set as $b^{-1}=2/3$. The optical depth can be
computed from the cross sections in a standard manner \cite{RJS96,RL03}.

In the present implementation, it is not necessary to
artificially divide the system into neutrino-trapped and 
free-streaming regions by the single neutrino-trapping density.
Therefore there is no discontinuous boundary
which existed in the previous leakage schemes \cite{EP81,vRL81,Kotake03}.

As the local production reactions of neutrinos,
the electron and positron captures\cite{FFN85} ($\gamma_{\nu e}^{\rm ec}$ and
$\gamma_{\bar{\nu} e}^{\rm pc}$),
the electron-positron pair annihilation\cite{CHB86}
($\gamma_{\nu_{e} \bar{\nu}_{e}}^{\rm pair}$ for electron-type neutrinos
and $\gamma_{\nu_{x} \bar{\nu}_{x}}^{\rm pair}$ for the other type),
the plasmon decays\cite{RJS96}
($\gamma_{\nu_{e} \bar{\nu}_{e}}^{\rm plas}$ and
$\gamma_{\nu_{x} \bar{\nu}_{x}}^{\rm plas}$),
and the Bremsstrahlung processes\cite{BRT06}
($\gamma_{\nu_{e} \bar{\nu}_{e}}^{\rm Brems}$ and
$\gamma_{\nu_{x} \bar{\nu}_{x}}^{\rm Brems}$) 
are considered in this paper.
Then, the local rates for the neutrino fractions are
\beqn
&& 
\gamma_{\nu e}^{\rm local} = \gamma_{\nu e}^{\rm ec} + 
\gamma_{\nu_{e} \bar{\nu}_{e}}^{\rm pair} + 
\gamma_{\nu_{e} \bar{\nu}_{e}}^{\rm plas} +
\gamma_{\nu_{e} \bar{\nu}_{e}}^{\rm Brems}, \label{gnlocal}\\
&& 
\gamma_{\bar{\nu} e}^{\rm local} = \gamma_{\bar{\nu} e}^{\rm pc} + 
\gamma_{\nu_{e} \bar{\nu}_{e}}^{\rm pair} + 
\gamma_{\nu_{e} \bar{\nu}_{e}}^{\rm plas} +
\gamma_{\nu_{e} \bar{\nu}_{e}}^{\rm Brems}, \label{galocal}\\
&& 
\gamma_{\nu x}^{\rm local} = 
\gamma_{\nu_{x} \bar{\nu}_{x}}^{\rm pair} + 
\gamma_{\nu_{x} \bar{\nu}_{x}}^{\rm plas} +
\gamma_{\nu_{x} \bar{\nu}_{x}}^{\rm Brems}. \label{gxlocal}
\eeqn
Similarly, the local neutrino energy emission rate $Q_{\nu}^{\rm local}$ is given
by
\beqn
Q_{\nu}^{\rm local} = Q_{\nu e}^{\rm ec} + Q_{\bar{\nu} e}^{\rm pc} 
&+& 2\,(Q_{\nu_{e} \bar{\nu}_{e}}^{\rm pair} + 
        Q_{\nu_{e} \bar{\nu}_{e}}^{\rm plas} +
        Q_{\nu_{e} \bar{\nu}_{e}}^{\rm Brems}) \nonumber \\
&+& 4\,(Q_{\nu_{x} \bar{\nu}_{x}}^{\rm pair} + 
        Q_{\nu_{x} \bar{\nu}_{x}}^{\rm plas} +
        Q_{\nu_{x} \bar{\nu}_{x}}^{\rm Brems})\ . \label{Qlocal}
\eeqn
The explicit forms of the local rates in
Eqs. (\ref{gnlocal})--(\ref{Qlocal}) will be found in 
Appendices A and B.

We follow the recent work by Rosswog and Liebend\"orfer\cite{RL03} for
the diffusive neutrino emission rates $\gamma_{\nu}^{\rm diff}$ and 
$Q_{\nu}^{\rm diff}$ in Eqs (\ref{Q_leak}) and (\ref{g_leak}).
The explicit forms of $\gamma_{\nu}^{\rm diff}$ and $Q_{\nu}^{\rm diff}$
are presented in Appendix C.


%
\section{Equation of state}\label{S_EOS}
In this section we summarize details of EOSs adopted in our current code.

\subsection{Baryons}
\label{EOS_Baryon}

In the present version of our code, 
we employ an EOS by Shen et al.\cite{Shen98}, which is derived by  
the relativistic mean field theory\cite{RMF} based on the relativistic 
Br\"uckner-Hartree-Fock theory\cite{RBHF}.
The so-called parameter set TM1\cite{RMF} is adopted to reproduce 
characteristic properties of heavy nuclei. 
The maximum mass of a cold spherical neutron star in this EOS 
is much larger than the canonical neutron star mass $\approx 1.4M_{\odot}$
as $\approx 2.2 M_{\odot}$\cite{Shen98}.
The framework of the relativistic mean field theory 
is extended with the Thomas-Fermi spherical cell model approximation 
to describe not only the homogeneous matter but also 
an inhomogeneous one.

The thermodynamical quantities of dense matter
at various sets of $(\rho, Y_{p}, T)$ are
calculated to construct the numerical data table for simulation. 
The table covers a wide range of density
$10^{5.1}$--$10^{15.4}$ g/cm$^{3}$, electron fraction $0.0$--$0.56$,
and temperature $0$--$100$ MeV, which are required for supernova
simulation.
It should be noted that the causality is guaranteed to be satisfied in
this framework, whereas the sound velocity
sometimes exceeds the velocity of the light in the non-relativistic
framework, e.g., in the EOS by Lattimer and Swesty\cite{LS91}. 
This is one of the benefits of the relativistic EOS.

Although we employ the nuclear EOS by Shen et al. 
in this work, it is easy to replace
the EOS. In the future we plan to implement other EOSs such as a
hyperonic matter EOS\cite{Ishizuka}.

Because the table of the original EOS by Shen et al. 
does not include the thermodynamical quantities of the leptons 
(electrons, positrons, and neutrinos if necessary) and
photons, one has to consistently include them to the table.

%
\subsection{Electrons and Positrons}
%
To consistently calculate the pressure and the internal energy of 
the electron and positron, the charge neutrality condition $Y_{p} = Y_{e}$ should 
be solved to determine the electron chemical potential
$\mu_{e}$ for each value of the baryon rest-mass density $\rho$ and
the temperature $T$ in the EOS table.
Namely, it is required to solve the equation
\beq 
n_{e}(\mu_{e},T) \equiv n_{-} - n_{+} = \frac{\rho Y_{e}}{m_{u}}
\label{n_to_mu}
\eeq
in terms of $\mu_{e}$ for given values of $\rho$, 
$T$, and $Y_{e}\ (= Y_{p})$.
Here, $m_{u} = 931.49432$ MeV is the atomic mass unit, 
and $n_{-}$ and $n_{+}$ are the total number densities
(i.e., including the electron-positron pair) of
the electrons and positrons, respectively. 

Assuming that the electrons obey the Fermi-Dirac distribution (which
is derived under the assumption of thermodynamic equilibrium), the
number density ($n_{-}$), the pressure ($P_{-}$), and the internal
energy density ($u_{-}$) of the electron are written as\cite{Cox}
\beqn
n_{-} &=& \frac{1}{\pi^{2} \hbar^{3}} \int_{0}^{\infty} 
\frac{p^{2}dp}
{\exp \left[ -\eta_{e} + \tilde{\epsilon}/k_{B} T\right] + 1}, \\
P_{-} &=& \frac{1}{\pi^{2} \hbar^{3}} \int_{0}^{\infty} 
\frac{p^{3}(\partial \tilde{\epsilon}/\partial p)dp}
{\exp \left[ -\eta_{e} + \tilde{\epsilon}/k_{B} T\right] + 1}, \\
u_{-} &=& \frac{1}{\pi^{2} \hbar^{3}}\int_{0}^{\infty} 
\frac{p^{2}\tilde{\epsilon} dp}
{\exp \left[ -\eta_{e} + \tilde{\epsilon}/k_{B} T \right] + 1}. 
\eeqn
Here  $\hbar$, $k_{B}$, and $\eta_{e} \equiv \mu_{e}/k_{B}T$ are the Planck's
constant, the Boltzmann's constant and the so-called degeneracy parameter.
$\tilde{\epsilon}(p) = \sqrt{m_{e}^{2}c^{4} + p^{2}} - m_{e}c^{2}$ is the
kinetic energy of a free electron.
If we choose the zero point of our energy scale for electrons at
$\tilde{\epsilon} = 0$, we have to assign a total energy of
$\tilde{\epsilon}_{+} = \sqrt{m_{e}^{2}c^{4} + p^{2}} + m_{e}c^{2}$ to a free
positron\cite{Cox}.
Thus the number density ($n_{+}$), the pressure ($P_{+}$), and the internal
energy density ($u_{+}$) of positrons are given by\cite{Cox}
\beqn
n_{+} &=& \frac{1}{\pi^{2} \hbar^{3}}\int_{0}^{\infty} 
\frac{p^{2}dp}{\exp 
\left[ -\eta_{+} + \tilde{\epsilon}_{+}/k_{B} T\right] + 1}, \\
P_{+} &=& \frac{1}{\pi^{2} \hbar^{3}}\int_{0}^{\infty} 
\frac{p^{3}(\partial \tilde{\epsilon}_{+}/\partial p)dp}
{\exp \left[ -\eta_{+} + \tilde{\epsilon}_{+}/k_{B} T\right] + 1}, \\
u_{+} &=& \frac{1}{\pi^{2} \hbar^{3}}\int_{0}^{\infty} 
\frac{p^{2}(\tilde{\epsilon}+2m_{e}c^{2})dp}
{\exp \left[ -\eta_{+} + \tilde{\epsilon}_{+}/k_{B} T\right] + 1},
\eeqn
where $\eta_{+} = -\eta_{e}$ is the degeneracy parameter of the positrons.

%
\subsection{Photons}
%
The pressure and the specific internal energy density of photons are
given by
\beqn
P_{r} = \frac{a_{r}T^{4}}{3},\ \ 
\varepsilon_{r} = \frac{a_{r}T^{4}}{\rho},
\eeqn
where $a_{r}$ is the radiation constant $a_{r} =
(\pi^{2}k_{B}^{4})/(15c^{3}\hbar^{3})$ 
and $c$ is the velocity of light.

\subsection{Trapped neutrinos}
In this paper, the trapped-neutrinos are assumed to interact 
sufficiently frequently with matter that be thermalized. Therefore
they are described as ideal Fermi gases with the matter temperature.
Then, from the neutrino fractions $Y_{\nu}$,
the chemical potentials of neutrinos are calculated by solving
\beq
Y_{\nu} = Y_{\nu}(\mu_{\nu}, T).
\eeq
Using the chemical potentials, $\mu_{\nu}$, and the matter temperature, 
the pressure and the internal energy of the trapped-neutrinos are 
calculated in the same manner as for electrons.

\subsection{The sound velocity}

In the high-resolution shock-capturing scheme for hydrodynamics,
we in general need to evaluate the sound velocity $c_{s}$,
\beq
c_{s}^{\,2} = \frac{1}{h}\left[ 
\left.\frac{\partial P}{\partial \rho}\right|_{\epsilon}
+\frac{P}{\rho}
\left.\frac{\partial P}{\partial \epsilon}\right|_{\rho}
\right]. \label{defcs}
\eeq
The derivatives of the pressure are calculated by
\beqn
\left.\frac{\partial P}{\partial \rho}\right|_{\epsilon}
&=&
\sum_{i=B,e,r,\nu}
\left[
  \left.\frac{\partial P_{i}}{\partial \rho}\right|_{T}
  -\left.\frac{\partial P_{i}}{\partial T   }\right|_{\rho}
  \left(
  \sum_{j=B,e,r,\nu}
  \left.\frac{\partial \epsilon_{j}}{\partial \rho}\right|_{T}
  \right)
  \left( \sum_{k=B,e,r,\nu}
  \left.\frac{\partial \epsilon_{k}}{\partial T}\right|_{\rho}
  \right)^{-1}  
\right], \label{Prho} \\
\left.\frac{\partial P}{\partial \epsilon}\right|_{\rho}
&=&
 \left(\sum_{i=B,e,r,\nu}
 \left.\frac{\partial P_{i}}{\partial T}\right|_{\rho}
 \right)
 \left(\sum_{j=B,e,r,\nu}
 \left.\frac{\partial \epsilon_{j}}{\partial T}\right|_{\rho}
 \right)^{-1} , \label{Peps}
\eeqn
where '$B$', '$e$', '$ph$' and '$\nu$' in the sum denote the baryon, electron,
photons, and neutrino quantities. 

The derivatives for the baryon parts are evaluated by taking a
finite difference of the table data. On the other hand,
the derivatives for the electron parts can be evaluated 
semi-analytically.
Because there are in general the phase transition regions
in an EOS table for baryons and moreover the EOS may contain 
some non-smooth spiky structures, careful treatments are necessary when 
evaluating the derivatives of thermodynamical quantities. 
In the present EOS table, the derivatives are carefully evaluated 
so that there are no spiky behaviors in the resulting sound velocities.

\section{Basic equations and Numerical methods} \label{S_Numerical}

\subsection{Einstein's equation and gauge conditions}

The standard variables in the 3+1 decomposition are 
the three-dimensional metric $\gamma_{ij}$ and 
the extrinsic curvature $K_{ij}$
on the three-dimensional hypersurface\cite{York79} defined by
\beqn
\gamma_{\mu\nu} &\equiv& g_{\mu\nu} + n_{\mu}n_{\nu}, \\
K_{\mu\nu} &\equiv& - \frac{1}{2} \bL _{n} \gamma_{\mu\nu}, 
\eeqn
where $g_{\mu\nu}$ is the spacetime metric,
 $n_{\mu}$ is the unit normal to a three-dimensional
hypersurface, and $\bL _{n}$ is the Lie derivative with
respect to the unit normal $n^{\mu}$. 
Then we can write the line element in the form
\beq
ds^{2} = - \alpha^{2} dt^{2} + \gamma_{ij}(dx^{i}+\beta^{i}dt)
(dx^{j}+\beta^{j}dt),
\eeq
where $\alpha$ and $\beta^{i}$ are the lapse function
and the shift vector which describe the gauge degree of freedom.

In the BSSN reformulation\cite{SN95,BS98}, 
the spatial metric $\gamma_{ij}$ is conformally decomposed as
$\gamma _{ij} = e^{ 4\phi}\tilde{\gamma}_{ij}$
where the condition $\det (\tilde{\gamma}_{ij}) = 1$ is imposed for the
conformal metric $\tilde{\gamma}_{ij}$. 
From this condition, the conformal factor is written as
$\phi = \frac{1}{12}\ln \gamma$ and $ \gamma \equiv \det(\gamma_{ij})$. 
The extrinsic curvature $K_{ij}$ is decomposed into the trace part 
$K$ and the traceless part $A_{ij}$ as
$K_{ij} = A_{ij} + (1/3)\gamma_{ij}K$ . 
The traceless part $A_{ij}$ is conformally
decomposed as $A_{ij} = e^{4\phi}\tilde{A}_{ij}$.
Thus the fundamental quantities for the evolution equation are now
split into 
$\phi, \tilde{\gamma}_{ij}$, $K$, and $\tilde{A}_{ij}$.
Furthermore, the auxiliary variable 
$ F_{i} \equiv \delta^{jk}\partial_{k}\tilde{\gamma}_{ij} $ 
is introduced in the BSSN reformulation\cite{SN95}.

The basic equations to be solved are
\begin{eqnarray}
&& \left(\partial_{t} - \beta^{k}\partial_{k} \right) \phi =
 \frac{1}{6}\left(-\alpha K + \partial_{k} \beta^{k} \right) ,
\label{phidevelopP}\\
&& \left(\partial_{t} - \beta^{k}\partial_{k} \right)\tilde{\gamma}_{ij} = 
 -2 \alpha \tilde{A}_{ij} + \tilde{\gamma}_{ik}\partial_{j}\beta^{k} +
 \tilde{\gamma}_{jk}\partial_{i}\beta^{k} -
 \frac{2}{3}\tilde{\gamma}_{ij} \partial_{k}\beta^{k}, 
\label{tilGamdevelopP}\\
&& \left(\partial_{t} - \beta^{k}\partial_{k} \right) K = 
 - D^{k}D_{k} \alpha  + \alpha \left[ \tilde{A}_{ij}\tilde{A}^{ij} +
  \frac{1}{3}K^{2} \right]
 + 4 \pi \alpha \left( \rho_{h} + S \right) , \label{trKdevelopP}\\
&& \left(\partial_{t} - \beta^{k}\partial_{k} \right)\tilde{A}_{ij} =
 e^{-4\phi}\left[ \alpha \left(R_{ij} - \frac{1}{3} e^{4\phi}
 \tilde{\gamma}_{ij} R \right) - \left(D_{i}D_{j} \alpha -\frac{1}{3}e^{4\phi}
 \tilde{\gamma}_{ij} D^{k}D_{k}\alpha  \right)\right] \nonumber \\
&& \ \ \ \ \ \ \ \ \ \ \ \ \ \ \ \ \ \ \ \ \ \ \ \  
 + \alpha \left( K \tilde{A}_{ij} - 2 \tilde{A}_{ik} \tilde{A}^{k}_{\ j} \right)
 + \tilde{A}_{ik}\partial_{j}\beta^{k} + \tilde{A}_{jk}\partial_{i}\beta^{k}
 - \frac{2}{3} \tilde{A}_{ij} \partial_{k}\beta^{k} \nonumber \\
&& \ \ \ \ \ \ \ \ \ \ \ \ \ \ \ \ \ \ \ \ \ \ \ \ 
 - 8 \pi \alpha \left( e^{-4\phi} S_{ij} 
 - \frac{1}{3}\tilde{\gamma}_{ij}S \right), \label{AdevelopP} \\
&&
\left( \partial_{t} - \beta^{k}\partial_{k} \right)F_{i} = 
  -16\pi \alpha j_{i} \nonumber \\
&& \ \ \ \ \ \ \ \ \ \ + 2\alpha 
   \left\{
   f^{kj}\partial_{j}\tilde{A}_{ik} + \tilde{A}_{ik} \partial_{j}f^{kj} 
   - \frac{1}{2}\tilde{A}^{jl}\partial_{i}h_{jl}
   + 6 \tilde{A}^{k}_{\ i}\partial_{k}\phi 
   - \frac{2}{3}\partial_{i}K
  \right\} \nonumber \\
&& \ \ \ \ \ \ \ \ \ \   
  + \delta^{jk}
  \left\{
   -2 \tilde{A}_{ij} \partial_{k} \alpha 
   + \left(\partial_{k}\beta^{l}\right)\partial_{l}h_{ij}
   \right. \nonumber \\
&& \ \ \ \ \ \ \ \ \ \  \ \ \ \ \ \ \ \ \ 
   \left. + \partial _{k} \left(
	\tilde{\gamma}_{il}\partial_{j}\beta^{l} 
        + \tilde{\gamma}_{jl}\partial_{i}\beta^{l}
        -\frac{2}{3}\tilde{\gamma}_{ij} \partial_{l}\beta^{l}
     \right) 
\right\},
\label{FdevelopP}
\end{eqnarray}
where $^{(3)}R$, $^{(3)}R_{ij}$, and $D_{i}$ are  
the Ricci scalar, the
Ricci tensor, and the covariant derivative associated with
three-dimensional metric $\gamma_{ij}$, respectively. 
The matter source terms,
$\rho_{h} \equiv (T^{\rm Total})^{\alpha \beta} n_{\alpha}n_{\beta}$, 
$j_{i} \equiv -(T^{\rm Total})^{\alpha \beta} \gamma_{i\alpha}n_{\beta}$, 
and $S_{ij} \equiv (T^{\rm Total})^{\alpha \beta} \gamma_{i\alpha}\gamma_{j \beta}$, 
are the projections of the stress-energy tensor with respect to $n^{\mu}$ and
$\gamma_{\mu\nu}$, and $S \equiv \gamma_{ij}S^{ij}$.

We assume the axial symmetry of the spacetime and the so-called 
Cartoon method\cite{Cartoon,Shibata03} is adopted to avoid 
problems around the coordinate singularities of the cylindrical 
coordinates.
Except for this, the numerical schemes for solving the Einstein's 
equation are essentially the same as those in Ref.~\citen{BNS1}.
We use 4th-order finite difference scheme in the spatial direction and
the 3rd-order Runge-Kutta scheme in the time integration.
The advection terms such as $\beta^{i}\partial_{i}\phi$ are evaluated
by a 4th-order upwind scheme.

As the gauge conditions for the lapse, 
we use the so-called $1+\log$ slicing\cite{AB01}:
\beq
(\partial_{t} - \bL _{\beta})\alpha = -2K \alpha.
\eeq
It is known that the $1+\log$ slicing enables to perform a long term
evolution of neutron stars as well as has strong singularity avoidance
properties in the black hole spacetime.

The shift vector is determined by solving the following dynamical 
equation\cite{DG}
\begin{equation}
\partial_{t}\beta^{k} = \tilde{\gamma}^{kl} (F_{l} + \Delta t
 \partial_{t} F_{l}).
\label{Dynbeta}
\end{equation}
Here the second term in the right-hand side is necessary for
numerical stability\cite{DG}.

\subsection{The hydrodynamic equations in leakage scheme}\label{BasicEq}

The basic equations for general relativistic hydrodynamics 
in our leakage scheme are the continuity equation, 
the lepton-number conservation equations, and
the local conservation equation of the energy-momentum. 
We assume the axial symmetry of the spacetime and the hydrodynamics
equations are solved in the cylindrical coordinates $(\varpi, \varphi,
z)$ where $\varpi = \sqrt{x^{2}+y^{2}}$.
In the axisymmetric case, the hydrodynamics equations should be
written in the cylindrical coordinate. On the other hand, in the Cartoon
method\cite{Cartoon,Shibata03}, Einstein's equation are solved in
the $y=0$ plane for which $x=\varpi$, $u_{\varpi} = u_{x}$,
$u_{\varphi} = x u_{y}$, and the other similar relations hold for
vector and tensor quantities. Taking into these facts, the hydrodynamic
equations may be written using the Cartesian coordinates replacing
$(\varpi, \varphi)$ by $(x,y)$.
In the following, we write down explicit forms of the equations for the
purpose of convenience.
Numerical tests for basic parts of the code of solving the hydrodynamics 
equations are extensively performed in Ref.~\citen{Shibata03}.
The equations are solved using the third-order
high-resolution central scheme of Kurganov and Tadmor\cite{KT00,PhD}.

\subsubsection{The Continuity and lepton-number conservation equations}

The continuity equation for the baryon rest mass is
\beq
\nabla_{\alpha}(\rho u^{\alpha}) = 0 \label{conti}.
\eeq
As fundamental variables for numerical simulations, the following
quantities are introduced:
$\rho_{\ast} \equiv \rho w e^{6\phi}$ and $v^{i} \equiv u^{i}/u^{t}$
where $ w \equiv \alpha u^{t}$.
Then, the continuity equation is written as
\beq
 \partial_{t}(\rho_{\ast}) 
 + \frac{1}{x}\partial_{x}(\rho_{\ast}v^{x})
 +            \partial_{z}(\rho_{\ast}v^{z}) = 0 .
\label{continuS} 
\eeq

Using the continuity equation, the lepton-number conservation 
equations (\ref{dYe}) -- (\ref{dYno}) are written as
\beq
  \partial_{t}(\rho_{\ast}Y_{L}) 
+ \frac{1}{x}\partial_{x}(\rho_{\ast}Y_{L}v^{x})
+            \partial_{z}(\rho_{\ast}Y_{L}v^{z}) = \rho_{*}\gamma_{L},
\label{e-Y} 
\eeq
where $Y_{L}$ and $\gamma_{L}$ are abbreviated expressions of the
lepton fractions and the source terms.

\subsubsection{Energy-momentum conservation}

As fundamental variables for numerical simulations, we define the 
quantities
$\hat{u}_{i} \equiv hu_{i}$ and 
$\hat{e} \equiv hw - P(\rho w)^{-1}$.
Then, the Euler
equation $\gamma_{i}^{\alpha} \nabla_{\beta}
T^{\beta}_{\ \alpha} = - \gamma_{i}^{\alpha} Q^{\rm leak}_{\alpha}$, 
and the energy equation 
$n^{\alpha}\nabla_{\beta}T^{\alpha}_{\beta}
=-n^{\alpha}Q^{\rm leak}_{\alpha}$ can be 
written as
\beqn
&&\!\!\!\! 
  \partial_{t}(\rho_{\ast} \hat{u}_{A})
+ \frac{1}{x}\partial_{x}
     \left[x \left\{
            \rho_{\ast} \hat{u}_{A} v^{x}
          + P\alpha e^{6\phi}\delta^{x}_{\ A} \right\}
      \right]  
+ \partial_{z}
     \left[
            \rho_{\ast} \hat{u}_{A} v^{z}
          + P\alpha e^{6\phi}\delta^{z}_{\ A}
      \right]  
\nonumber \\
&&\!\!\!\!  \ \ \ \ \ \ \ \ \ \ 
=  
- \rho_{\ast}\left[ 
     w h \partial_{A}\alpha - \hat{u}_{i}\partial_{A}\beta^{i}
      + \frac{\alpha e^{-4\phi}}{2wh}
        \hat{u}_{k}\hat{u}_{l}\partial_{A}\tilde{\gamma}^{kl}
      - \frac{2\alpha h (w^{2} -1)}{w}
        \partial_{A}\phi \;
   \right] \nonumber \\ 
&&\!\!\!\!  \ \ \ \ \ \ \ \ \ \ \ \ \ 
+ P\partial_{A}(\alpha e^{6\phi}) 
+ \frac{(\rho_{*}u_{y}v^{y} + P\alpha e^{6\phi}) \delta^{x}_{A}}{x}  
- \alpha e^{6\phi} Q^{\rm leak}_{A}, \label{momS}  \\
&&\!\!\!\! 
  \partial_{t}\left( \rho_{\ast}\hat{u}_{y}\right)
+ \frac{1}{x^{2}}
   \partial_{x}\left( 
       x^{2} \rho_{\ast}\hat{u}_{y}v^{y}  \right)
+ \partial_{z}\left( 
             \rho_{\ast}\hat{u}_{y}v^{z}  \right)
=   -\alpha e^{6\phi} Q^{\rm leak}_{y}, \\
&&\!\!\!\! 
  \partial_{t}(\rho_{\ast}\hat{e})  
+ \frac{1}{x}
   \partial_{x}\left[ x \left\{
        \rho_{\ast}v^{x}\hat{e}
      + P e^{6\phi}(v^{x}+\beta^{x}) \right\}
  \right]
+ \partial_{z}\left[
        \rho_{\ast}v^{z}\hat{e}
      + P e^{6\phi}(v^{z}+\beta^{z})
  \right]
\nonumber \\
&&\!\!\!\!  \ \ \ \ \ \ \ \ \ \ \ 
= \alpha e^{6\phi} PK + 
\frac{\rho_{\ast}}{u^{t}h} \hat{u}_{k}\hat{u}_{l}K^{kl}
 - \rho_{\ast}\hat{u}_{i}\gamma^{ij}D_{j}\alpha 
 - \alpha e^{6\phi} Q^{\rm leak}_{\alpha}n^{\alpha},  \label{eneS} 
\eeqn
where the subscript $A$ denotes $x$ or $z$ component.

The evolution equation of streaming-neutrinos 
$\nabla_{\beta}(T^{\nu,{\rm S}})^{\beta}_{\alpha} = Q^{\rm leak}_{\alpha}$ gives
\beqn
&&\!\!\!\!\!\!\!\!
  \partial_{t}(\hat{E}) 
+ \frac{1}{x}\partial_{x}\left[
    x (\alpha \hat{F}^{x} - \beta^{x}\hat{E}) \right]  
+ \partial_{z}\left[
      (\alpha \hat{F}^{z} - \beta^{z}\hat{E}) \right]  
  \nonumber \\
&&\!\!  \ \ \ \ \ \ \ \    
  = \frac{\alpha \hat{E} K}{3} - \hat{F}^{k}\partial_{k}\alpha 
   +\alpha e^{6\phi} Q^{\rm leak}_{a}n^{a} , \\
&&\!\!\!\!\!\!\!\!
  \partial_{t}(\hat{F}_{A}) 
+ \frac{1}{x} \partial_{x}\left[ x
   \left(\frac{1}{3}\alpha \hat{E} \delta^{x}_{A} -
   \beta^{x}\hat{F}_{A} \right) \right]
+ \partial_{z}\left[
   \left(\frac{1}{3}\alpha \hat{E} \delta^{z}_{A}-\beta^{z}\hat{F}_{A} \right)\right]
  \nonumber \\
&&\!\!  \ \ \ \ \ \ \ \ 
  = -\hat{E}\partial_{A}\alpha + \hat{F}_{k}\partial_{A}\beta^{k} 
    + 2 \alpha \hat{E}\partial_{A}\phi 
    + \frac{(\hat{E}/3 - \hat{F}_{y}\beta^{y})\delta^{x}_{A}}{x} 
    + \alpha e^{6\phi} Q^{\rm leak}_{A}, \\
&&\!\!\!\!\!\!\!\!
  \partial_{t}(\hat{F}_{y}) 
- \frac{1}{x^{2}} \partial_{x}\left[ x^{2}
   \beta^{x}\hat{F}_{y}  \right]
- \partial_{z}\left[
   \beta^{z}\hat{F}_{y} \right] = \alpha e^{6\phi} Q^{\rm leak}_{y},
\eeqn
where $\hat{E} = e^{6\phi}E$ and $\hat{F}_{i} = e^{6\phi}F_{i}$, and the 
subscript $A$ again denotes $x$ or $z$ component.
The closure relation $P_{\alpha \beta} = E\gamma_{\alpha \beta}/3$ is
also substituted.

\subsection{Recover of ($\rho$, $Y_{e}$/$Y_{l}$, $T$)} \label{Reconst}

The quantities numerically evolved in the relativistic hydrodynamics
are the conserved quantities ($\rho_{*}$, $\hat{u}_{i}$, $\hat{e}$)
and the lepton fraction $Y_{e}$ or $Y_{l}$. 
The argument variables,
($\rho$, ($Y_{e}$ or $Y_{l}$), $T$), of the EOS table, 
together with the weight factor $w = \sqrt{1+\gamma^{ij}u_{i}u_{j}}$, 
should be calculated from the conserved
quantities at each time slice. 
Note that the electron ($Y_{e}$) or lepton fraction ($Y_{l}$) is readily
given by numerical evolution 
at each time slice whereas $\rho$, $u_{i}$, and $T$ are not.
This fact requires us to find an efficient method for determining $w$.

\subsubsection{Non-$\beta$-equilibrium case}
In the case that the $\beta$-equilibrium condition is not satisfied,
the argument quantities ($\rho$, $Y_{e}$, $T$) can be reconstructed
from the conserved quantities in the following straightforward manner.

\begin{enumerate}
\item Give a trial value of $w$, referred to as $\tilde{w}$. 
Then, one obtains a trial value of the rest mass density from
$\tilde{\rho} = \rho_{*}/(\tilde{w} e^{6\phi})$.
\item A trial value of the temperature, $\tilde{T}$, can be
obtained by solving the following equation:
\beq
\hat{e} = \left(1+\tilde{\varepsilon} + 
\frac{\tilde{P}}{\tilde{\rho}}\right)\tilde{w} -
\frac{\tilde{P}}{\tilde{\rho}\tilde{w}}
\equiv \tilde{e}(\tilde{\rho}, Y_{e}, \tilde{T}).
\eeq 
Here, one dimensional search over the EOS table is
required to obtain $\tilde{T}$.
\item The next trial value of $w$ is given by 
$\tilde{w} =
      \sqrt{1+e^{-4\phi}\tilde{\gamma}^{ij}\hat{u}_{i}\hat{u}_{j}\tilde{h}^{-2}}$, 
where the specific enthalpy was calculated
as $\tilde{h} = \tilde{h}(\tilde{\rho}, Y_{e}, \tilde{T})$ in the step 2.
\item Repeat the procedures (1)--(3) until a required degree of convergence
  is achieved. Convergent solutions of the temperature and $w$ are
  obtained typically in 10 iterations.
\end{enumerate}

\subsubsection{The $\beta$-equilibrium case}
On the other hand,
in the case that the $\beta$-equilibrium condition is satisfied,
one has to reconstruct the argument quantities ($\rho, Y_{e}, T$) from 
the conserved quantities and $Y_{l}$, under the assumption of the 
$\beta$-equilibrium. In this case, two-dimensional recover
$(Y_{l}, \hat{e}) \  \Longrightarrow \ (Y_{e}, T)$
would be required for a given value of $\tilde{w}$. A serious problem is 
that in this case, there may be more than one combination 
of ($Y_{e}$, $T$) which gives the same values of $Y_{l}$ and $\hat{e}$.
Therefore, we have to adopt a different method 
to recover ($\rho, Y_{e}, T$).
Under the assumption of the $\beta$-equilibrium, 
the electron fraction is related to the total lepton fraction: 
$Y_{e} = Y_{e}(\rho, Y_{l}, T)$. Using this relation, the EOS table
can be rewritten in terms of the argument variables of 
($\rho$, $Y_{l}$, $T$). 
Then, the same strategy as in the non-$\beta$-equilibrium
case can be adopted. Namely,
\begin{enumerate}
\item Give a trial value $\tilde{w}$. 
Then one obtains a trial value of the rest mass density.
\item A trial value of the temperature can be
obtained by solving 
$ \hat{e} = \tilde{e}(\tilde{\rho}, Y_{l}, \tilde{T})$,
with one dimensional search over the EOS table.
\item The next trial value of $w$ is given by 
$\tilde{w} = \sqrt{1+e^{-4\phi}\tilde{\gamma}^{ij}\hat{u}_{i}\hat{u}_{j}\tilde{h}^{-2}}$. 
\item Repeat the procedures (1)--(3) until a required degree of convergence
  is achieved. The electron fraction is given as 
  $Y_{e} = Y_{e}(\rho, Y_{l}, T)$ in the (new) EOS table.
\end{enumerate}

In the case of a simplified or analytic EOS, 
the Newton-Raphson method may be applied to recover the primitive
variables. In the case of a tabulated EOS, by contrast,
the Newton-Raphson method may not be a good approach because it requires
derivatives of thermodynamical quantities which in general cannot be 
calculated precisely from a tabulated EOS by the finite differentiating 
method.

\subsection{Grid Setting}\label{Grid}

In numerical simulations, 
we adopt a nonuniform grid, in which 
the grid spacing is increased as
\beq
d x_{j+1} = (1 + \delta) d x_{j}, \ \ \ \ d z_{l+1} = (1 + \delta) d z_{l}
\eeq
where $d x_{j} \equiv x_{j+1} - x_{j}$, $d z_{l} \equiv z_{l+1} - z_{l}$, 
and $\delta$ is a constant.
In addition, a regridding technique\cite{Shibata02,Sekiguchi05} is adopted
to assign a sufficiently large number of grid points inside the
collapsing core, saving the CPU time efficiently.
The regridding is carried out whenever the characteristic radius
of the collapsing star decreases by a factor of a 2--3. 
At each regridding, the minimum grid spacing is decreased by a factor
of $\sim 2$ while the geometrical factor $\delta$ is unchanged 
(see Table \ref{regrid}).

All the quantities on the new grid are calculated
using the fifth-order Lagrange interpolation. 
To avoid discarding the matter in the outer region, we also increase the 
grid number at each regridding.
For the regridding, we define 
a relativistic gravitational potential
$\Phi_c \equiv 1 -\alpha_c~ (\Phi_c>0)$ where $\alpha_c$ is the central value
of the lapse function.
Because $\Phi_c$ is approximately proportional to $M/R$ where $M$ and $R$
are characteristic mass and radius of the core, 
$\Phi_c^{-1}$ can be used as a measure of the characteristic
length scale of the stellar core for the regridding. 
In Table \ref{regrid}, we summarize the regridding parameters of each
level of the grid.

\begin{table}[t]
 \begin{center}
  \begin{tabular}{c|c|ccccc} \hline
   Model &  & $\Phi_{c} \le 0.0125 $ & $  \le \Phi_{c} \le 0.025 $
           & $  \le \Phi_{c} \le 0.05 $ 
           & $  \le \Phi_{c} \le 0.1 $ & $\Phi_{c} \ge 0.1$ \\
           \hline
  S15    & $\Delta x_{0}$ & 3.26 & 1.60 & 0.820 & 0.414 & 0.217 \\
         & $\delta$  & 0.002 & 0.002 & 0.002 & 0.002 & 0.002  \\
         & $N$     & 444    & 668    & 924    & 1212   & 1532    \\
         & $L$ (km)& 2330   & 2239   & 2188   & 2124   & 2103    \\ \hline
  S15    & $\Delta x_{0}$ & 5.10 & 2.90 & 1.44 & 0.760 & 0.396 \\
 (low)   & $\delta$  & 0.002 & 0.00215 & 0.0023 & 0.00245 & 0.0026  \\
         & $N$     & 316 & 444 & 636 & 828 & 1020  \\
         & $L$ (km)& 2244 & 2151 & 2073 & 2043 & 2000 \\ \hline
  \end{tabular}
 \end{center}
\caption{Summary of the regridding procedure. The values of the minimum
 grid spacing $\Delta x_{0}$ (in units of km), 
 the non-uniform-grid factor $\delta$, and
 the grid number $N$ for each range of $\Phi_{c} = 1 -\alpha_c$ are
 listed.}\label{regrid}
\end{table}

\section{Results} 
\label{S_Results}

As a test problem, we perform a collapse simulation of 
spherical presupernova core and compare the results with 
those in the previous studies, to see the validity of the 
present code.
Most of the following results are not novel astrophysically, 
but are novel in the sense that stellar core collapse can be 
followed by a {\it multidimensional fully general relativistic 
simulation taking account of microphysical processes}.
In \S~\ref{bounce_shock}, \S~\ref{shock_stall}, and 
\S~\ref{convective_activities}, we first review the basic features of 
the collapse dynamics and the shock formation, the stall of shock, 
and convective activities. Then we compare our results with those 
in the previous studies in \S~\ref{comparison}.

\subsection{Initial condition}\label{init_con}

In this paper, we adopt a recent presupernova model of massive
star by Woosley, Heger, and Weaver\cite{WHW02}: 
$15M_{\odot}$ model with solar metallicity (hereafter S15 model).
We follow the dynamical evolution of the central part which 
constitutes the Fe core and the inner part of the Si-shell. 
We read in the density, the electron fraction, the temperature and the
velocity ($v_{i}$) of the original initial data and derive other 
thermodynamical quantities using the EOS table.

Note that the procedure of remapping the original initial data into the
grid adopted in the numerical simulations is coordinate-dependent in 
general relativity. In this paper, we read in the original data {\it as
a function of the coordinate radius}. In this case, the baryon rest-mass
of the core is slightly larger than the original one, because
it is defined by
\beq
M_{*} = \int \rho_{*} dx^{3} = \int \rho (w e^{6\phi}) d^{3}x,
\label{restM}
\eeq
where $w e^{6\phi} > 1$.

\begin{figure}[t]
  \begin{center}
    \includegraphics[scale=1.0]{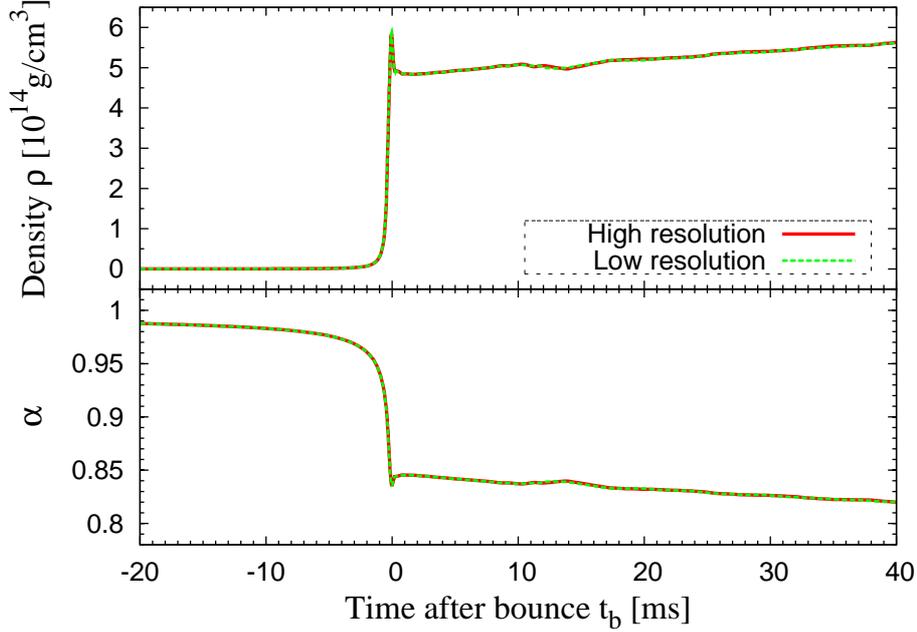} 
  \end{center}
  \caption{Evolution of the central density $\rho_{c}$ (upper panel)
    and the central value of the lapse function $\alpha_{c}$ (lower
    panel). The solid curves are results for the finer
    grid resolution and the dotted curves are results of the coarser
    grid resolution.}\label{rho-alp}
\end{figure}

\subsection{Core bounce and shock formation}\label{bounce_shock}

Figure \ref{rho-alp} displays the time evolution of
the central rest-mass density $\rho$ and the central value of the
lapse function. 
This figure shows that the stellar core collapse to a neutron
star can be divided into three phases; the infall phase, the bounce
phase, and the quasi-static evolution phase 
(see Refs.~\citen{Monch91} and \citen{Zwerg97} for the case of rotational collapse).
The general feature of the collapse is as follows.

The infall phase sets in due to gravitational instability
of the iron core triggered by the sudden softening of the EOS,
which is associated primarily with the electron capture and 
partially with the photo-dissociation of the heavy nuclei.
The collapse in an early phase proceeds almost homologously.
However, the collapse in the central region is accelerated 
with time because the electron capture reduces the degenerate pressure
of electrons which provides the main part of the total pressure.
Furthermore, the neutrino emission associated with the electron
capture reduces the thermal pressure of the core. 
Here the inner part of the core, which collapses nearly
homologously with a subsonic infall velocity, constitutes the inner
core. On the other hand, the outer region in which the infall
velocity is supersonic constitutes the outer core.

The collapse proceeds until the central part of the iron core reaches
the nuclear density ($\sim 2\times 10^{14}$ g/cm$^{3}$), and then, 
the inner core experiences the bounce. 
Because of its large inertia and large kinetic energy induced by the
infall, the inner core overshoots its hypothetical equilibrium state. 
The stored internal energy of the inner
core at the maximum compression is released through 
strong pressure waves generated inside the inner core.
The pressure waves propagate from the center to the outer region 
until they reach the sonic point located at the edge of the inner
core. Because the sound cones tilt inward beyond the sonic point, the
pressure disturbance cannot propagate further and forms a shock
just inside the sonic point. 
A shock wave is formed at the edge of the inner core and
propagates outward. 

After this phase, the proto-neutron star experiences the quasi-static
evolution phase. In this phase, the central value of density (the lapse
function) increases (decreases) gradually, because the matter in the
outer region falls into the proto-neutron star and because neutrinos are
emitted carrying away the energy and lepton-number from the
proto-neutron star.

\begin{figure}[t]
  \begin{center}
      \includegraphics[scale=1.0]{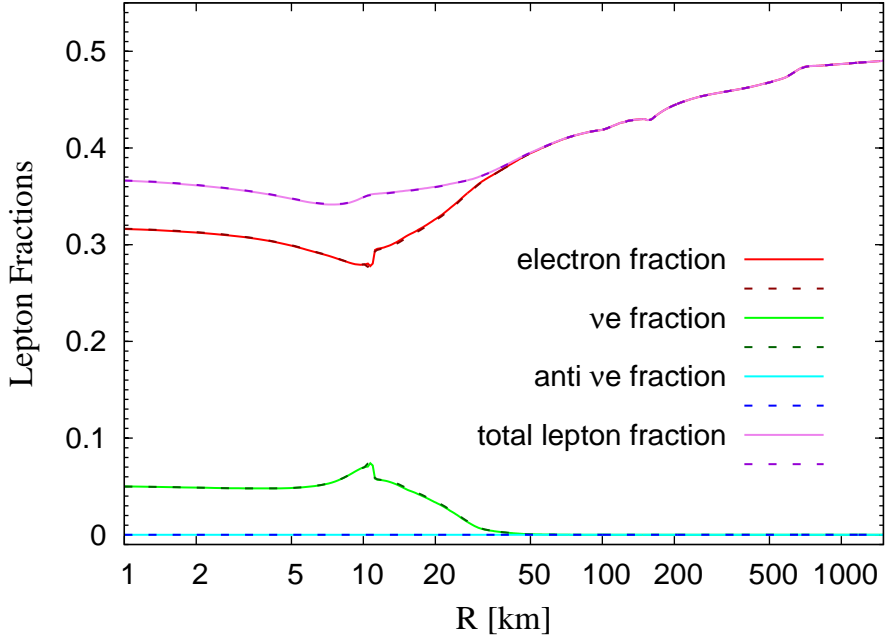}
  \end{center}
  \caption{The radial profiles of the electron, $\nu_{e}$-, $\bar{\nu}_{e}$-, 
    and the total lepton fraction at the bounce. 
    The results for the finer grid resolution (solid curve) 
    and for the coarser grid resolution 
    (the dotted curves) are shown together. The two results are almost 
    identical.
  }\label{lepb}
\end{figure}

Figure \ref{lepb} shows the radial profiles in the equator of the lepton 
fractions at the bounce. The central values of the electron, the
electron-neutrinos, and the total-lepton fractions are $\approx 0.32$,
0.05, and 0.37, respectively. The electron-anti-neutrino fraction is
almost zero through out the core because only very small amount of
positrons exist due to the high degree of electron degeneracy. 

\subsection{Neutrino bursts and stall of shock}\label{shock_stall}

\begin{figure}[t]
  \begin{center}
    \includegraphics[scale=1.0]{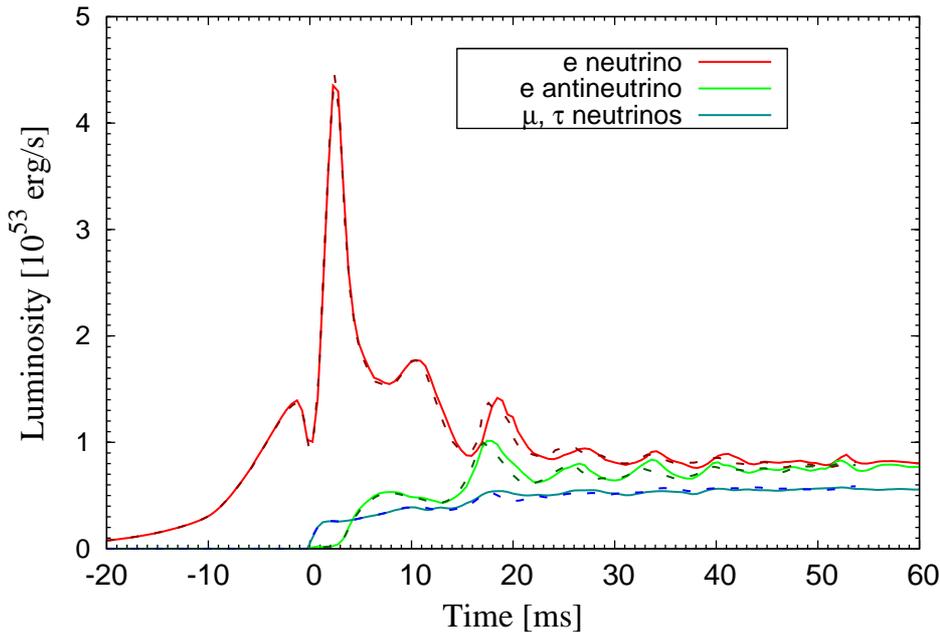}
  \end{center}
  \caption{Time evolution of the neutrino luminosities. 
    The results in the finer grid resolution (solid curves) and 
    in the coarser grid resolution (dashed curves) are shown together. 
    The two results are approximately identical until the convective phase sets in,
    whereas small disagreement is found in the convective phase.}\label{nlum}
\end{figure}

\begin{figure}[t]
  \begin{center}
      \includegraphics[scale=1.1]{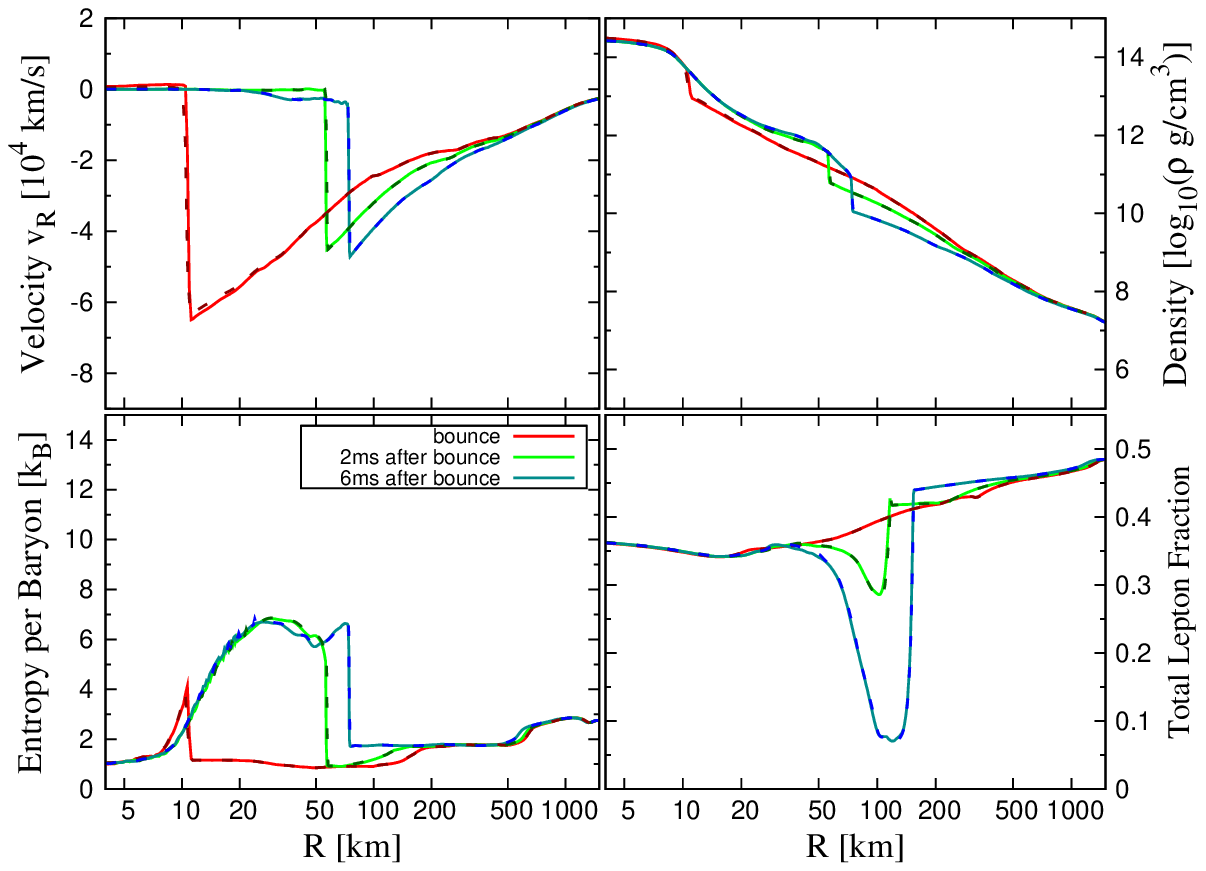}
  \end{center}
  \caption{The radial profiles of the infall velocity, the density, 
    the entropy per baryon, and the total lepton fraction at bounce,
    at 2 ms and 6ms after bounce. 
    The results for the finer grid resolution (solid curves) 
    and for the coarser grid resolution 
    (the dotted curves) are shown together and they are shown to be
    approximately identical.
}\label{qalx}
\end{figure}

As the shock wave propagates outward, the kinetic energy of the infall
matter is converted into the thermal energy behind the
shock.
The conversion rate of infall kinetic energy may be estimated 
approximately as
\beqn
L_{\rm heat} 
&\sim& 4\pi R_{s}^{2} (\rho_{\rm infall} v_{\rm infall}^{3}/2) 
\nonumber \\
&\sim& 
1.4 \times 10^{53}\, {\rm ergs/s}\,
\left(\frac{R_{s}}{100\,{\rm km}}\right)^{2}
\left(\frac{\rho_{\rm infall}}{10^{9}\,{\rm g/cm}^{3}}\right)
\left(\frac{v_{\rm infall}}{0.2 c}\right)^{3}, 
\label{shockp}
\eeqn
where $R_{s}$ and $\rho_{\rm infall}$ are radius of the shock wave and
the density of infall matter, and we here recover the velocity of the
light ($c$).
Here, we assume that all the kinetic energy is converted 
to the thermal energy.

At the same time, the shock wave suffers from the energy loss by the
photo-dissociation of the iron to $\alpha$-particles and free nucleons. 
The fraction of this energy loss is\cite{ST83}
\beq
\epsilon_{\rm diss} \sim 1.5 \times 10^{51} {\rm ~ergs ~per~} 0.1M_{\odot}.
\eeq
Thus, the energy loss rate due to the photo-dissociation is 
\beq
L_{\rm diss} \sim \dot{M}_{\rm shock} \epsilon_{\rm diss} 
\sim 1.1 \times 10^{53}\, {\rm ergs/s}\,
\left(\frac{R_{s}}{100\,{\rm km}}\right)^{2}
\left(\frac{\rho_{\rm infall}}{10^{9}\,{\rm g/cm}^{3}}\right)
\left(\frac{v_{\rm infall}}{0.2 c}\right),
\label{dissp}
\eeq
where 
$\dot{M}_{\rm shock} \sim 4\pi R_{s}^{2}\rho_{\rm infall}v_{\rm infall}$ 
is mass-accretion rate to the shock front.

The ratio of $L_{\rm heat}$ to $L_{\rm diss}$ is 
\beq
\frac{L_{\rm heat}}{L_{\rm diss}} \approx 1.2 
\left( \frac{v_{\rm infall}}{0.2c} \right)^{2}.
\eeq
Therefore the energy loss rate by the photo-dissociation will eventually
overcome the hydrodynamic power, because the infall velocity, which is 
$\approx (GM_{\rm ic}/R_{s})^{1/2}$, decreases as
the shock wave propagates outward.

Furthermore, when the shock wave crosses the neutrino-sphere, 
spiky burst emissions of neutrinos, the so-called neutrino bursts, occur:
Neutrinos in the hot post-shock region are copiously emitted without
interacting core matter.
Figure \ref{nlum} shows the neutrino luminosity as a function of time 
calculated by\cite{PhD,SSR07}
\beq
L_{\nu} = \int \alpha e^{6\phi} u_{t} \dot{Q}_{\nu} d^{3}x .
\eeq
The peak luminosity is $L_{\nu _{e}} \approx 4.5 \times 10^{53}$ ergs/s.
This neutrino burst significantly
reduces the thermal energy of the shock. Consequently, 
the shock wave stalls at $\approx 80$ km soon after the neutrino burst.
The peak luminosity and the shock-stall radius agree approximately with the 
previous one-dimensional fully general relativistic study\cite{Lieb05b}.

When the shock wave stalls, negative gradients of the entropy per baryon 
and the total-lepton (electron) fraction appear because neutrinos 
carry away both the energy and the lepton number.
Figure \ref{qalx} shows the radial profiles of the infall velocity, 
the density, the entropy per baryon, and the total lepton fraction in the
equator.
This figure clearly shows that negative gradients of the entropy per 
baryon and the total lepton fraction are formed above the neutrino sphere. 
As we shall see in \S~\ref{convective_activities}, 
such configurations are known to be unstable to convection, 
which is known to as the proto-neutron star convection. 

\subsection{Convective activities}\label{convective_activities}


\begin{figure}[t]
  \begin{center}
    \includegraphics[scale=1.1]{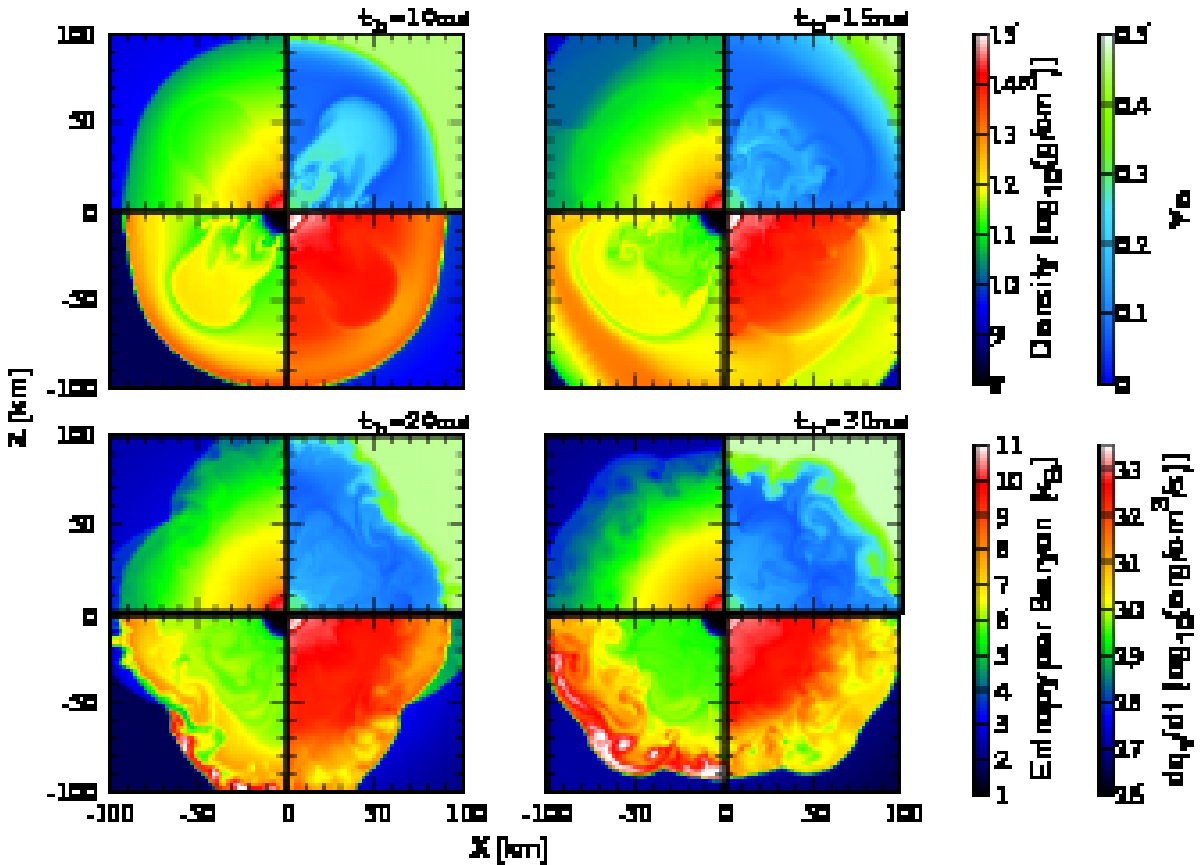}
  \end{center}
  \caption{Snapshots of the contours of the density (top left panels),  
    the electron fraction $Y_{e}$ (top right panels), the entropy per baryon
    (bottom left panels), and the local neutrino energy emission rate 
    (bottom right panels) in the $x$-$z$ plane at selected time slices.  
}\label{con}
\end{figure}

\begin{figure}[t]
  \begin{center}
    \includegraphics[scale=1.2]{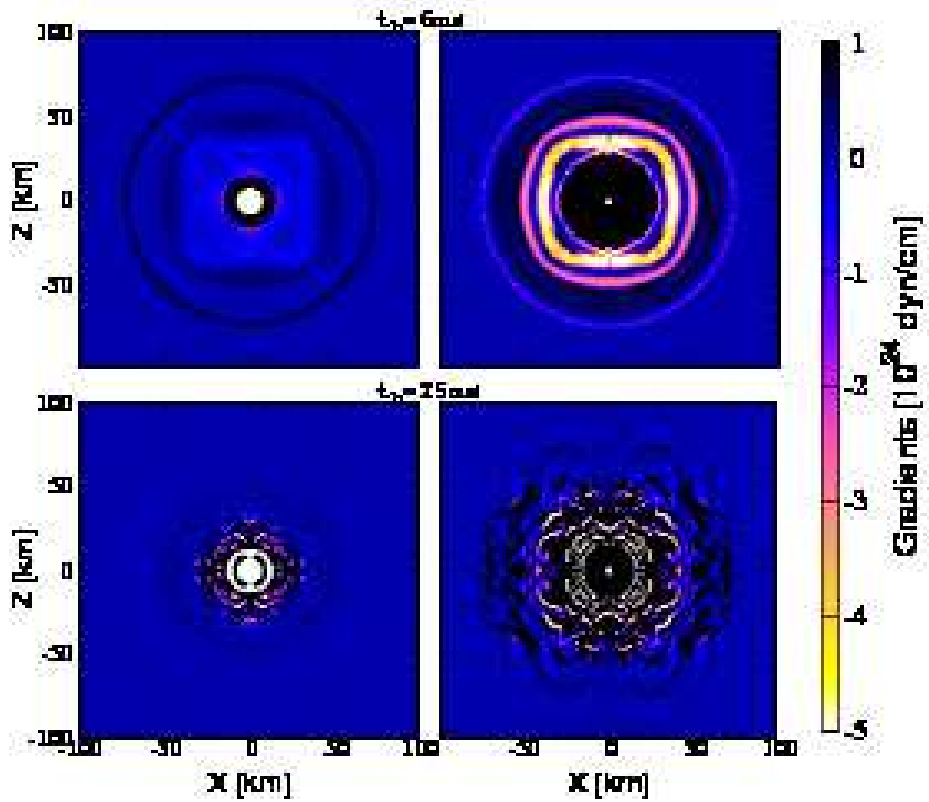}
  \end{center}
  \caption{Snapshots of the contours of gradients associated with
    the entropy per baryon
    $(\partial P / \partial s)_{Y_{l},\rho} (ds/dr) $ (right panels) 
    and associated with the lepton fraction
    $(\partial P / \partial Y_{l})_{s,\rho} (dY_{l}/dr) $ (left panels)
    in the $x$-$z$ plane at selected time slices.  
}\label{convG}
\end{figure}


Let us investigate the stability of the envelope of the proto-neutron
star following Lattimer and Mazurek\cite{LM81}.
We consider the following parameter
\beq
N^{2} \equiv \frac{g_{\rm eff}}{\rho} \left[
 \left(\frac{d \rho}{d r}\right)_{\rm amb} 
-\left(\frac{d \rho}{d r}\right)_{\rm blob} 
\right],
\label{VB0}
\eeq 
where $g_{\rm eff}$ is the effective gravitational acceleration defined 
to be positive in
the negative radial direction, the subscript 'amb' refers to the ambient
core structure, and 'blob' denotes the blob element which is under an
isolated displacement. The condition $N^{2} <0$
implies that the structure is unstable to convective overturn
(e.g. Ref.~\citen{LM81}). 

Assuming that the fluid elements maintain the pressure equilibrium with
its surroundings, we have 
\beq
\left(\frac{d \rho}{d r}\right)_{\rm blob} =
\left(\frac{d \rho}{d P}\right)_{\rm blob}
\left(\frac{dP}{dr}\right)_{\rm amb}.
\eeq
Using this relation, Eq. (\ref{VB0}) is written as
\beq
N^{2} =  \frac{g_{\rm eff}}{\rho} 
\left(\frac{d \rho}{d P}\right)_{\rm blob}
\left[
 \left(\frac{d P}{d \rho}\right)_{\rm blob}
 \left(\frac{d \rho}{d r}\right)_{\rm amb} 
-\left(\frac{d P}{d r}\right)_{\rm amb} 
\right].
\eeq
Because the pressure is a function of the entropy per baryon, 
the density, and the lepton fraction, $(d P/d r)_{\rm amb}$ 
is rewritten to give\cite{LM81}
\beq
N^{2} = \frac{g_{\rm eff}}{\rho}
\left(\frac{\partial \rho}{\partial P}\right)_{s, Y_{l}}
\left[
 \left(\frac{\partial P}{\partial s}\right)_{\rho, Y_{l}}
 \left(\frac{d s}{d r}\right)_{\rm amb}
+\left(\frac{\partial P}{\partial Y_{l}}\right)_{\rho, s}
 \left(\frac{d Y_{l}}{d r}\right)_{\rm amb}
\right] \label{VB1}.
\eeq
Here, we also assume that the blob elements do not 
interact the ambient matters both thermally and chemically, 
i.e. $ds = dY_{l} = 0$ for the blob. Then, we have
\beq
\left(\frac{dP}{d\rho}\right)_{\rm blob} = 
\left(\frac{\partial P}{\partial \rho} \right)_{s,Y_{l}}.
\eeq

Equation (\ref{VB1}) shows that when the pressure derivatives of given EOS 
($(\partial  P/\partial  s)_{\rho Y_{e}}$ and
 $(\partial  P/\partial  Y_{l})_{\rho s}$) are
positive, configurations with negative
gradients of entropy and $Y_{l}$ ($N^{2} < 0$) are unstable. 
(Note that in the above treatment, we have ignored the
dissociative effects caused by energy and lepton transports due to
neutrinos.)
Thus, the negative gradients of the entropy per 
baryon and the total lepton fraction formed
above the neutrino sphere lead to the convective overturn (the
proto-neutron star convection).  
Indeed, convection occurs in our simulation.

Figure \ref{con} shows contours of the density, 
the electron fraction, the entropy per baryon,
and the neutrino energy-emission rate.
Convective motions are
activated at about 8 ms after the bounce in the region located above the
neutrino-sphere where the gradients of the entropy per baryon 
and $Y_{l}$ are imprinted (see Fig. \ref{qalx}). 
At about 10 ms after the bounce, the lepton rich, hot blobs rise to form
'fingers' (see in top left panel in Fig. \ref{con}). 
Note that the neutrino energy emission rate in this finger is 
relatively higher than that in other region. 
This is responsible for the small hump seen in the
time-evolution of neutrino luminosity (see Fig. \ref{nlum}). 
Subsequently, the hot fingers expand to form 'mushroom structures', and
push the surface of the stalled shock 
(see top right panel in Fig. \ref{con}). 
At the same time, the lepton poor, colder matters sink down to 
the proto-neutron star ($r\lesssim 20$ km). 
The entropy per baryon just behind the shock increases to be 
$s \gtrsim 10 k_{B}$ and the stalled shock gradually moves outward
to reach $r\approx 200$ km. 
As the hot, lepton rich matters are dug out from the
region below the neutrino-sphere, 
the neutrino luminosity is enhanced (see Fig. \ref{nlum}). 
However, the energy released in the convective overturn is not
sufficient to keep pushing the shock wave, and eventually, the shock
stalls and turns to be a standing accretion shock 
(bottom two panels of Fig. \ref{con}). 
All these features qualitatively agree with the previous
multidimensional Newtonian
simulations\cite{BF93,Herant94,BHF95,KJM96,MJ97}.
A more detailed comparison with the previous simulations is given in
\S~\ref{comparison}.

It will be interesting to investigate which gradient 
(entropy per baryon or electron fraction) is more responsible for 
the convection. To see this, we calculate the gradients 
associated with the entropy per baryon 
$(\partial P / \partial s)_{Y_{l},\rho} (ds/dr) $ 
(right panels in Fig. \ref{convG}) 
and associated with the lepton fraction
$(\partial P / \partial Y_{l})_{s,\rho} (dY_{l}/dr)$ 
(left panels in Fig. \ref{convG}).
This figure clearly shows that negative gradient of the entropy 
per baryon is more important for the convection activated promptly 
after the bounce.

\subsection{Comparison with the previous studies}\label{comparison}

To check the validity of the code, the results presented in
\S~\ref{bounce_shock}, \S~\ref{shock_stall}, and
\S~\ref{convective_activities} are compared with the previous simulations.

\subsubsection{Comparison of the results before the convection sets in}
We first compare our results with those in the state-of-the-art
one-dimensional (1D) simulations in full general 
relativity\cite{Lieben01,Lieb04,Lieb05b,Sumi05}, 
in which 1D general relativistic Boltzmann equation is 
solved for neutrino transfer with relevant weak interaction processes.
Because neutrino heating processes ($\nu_{e}+n \rightarrow p+e^{-}$ and 
$\bar{\nu}_{e}+p\rightarrow n+e^{+}$) are not included in the present 
implementation, and on the other hand, multidimensional effects such as
convection cannot be followed in the one-dimensional reference simulations, 
we pay particular attention to comparing results during the collapse and 
the early phase ($\sim 10$ ms) after the bounce (see results in 
\S~\ref{bounce_shock} and \S~\ref{shock_stall}). 

Our radial profiles of the lepton fractions at the bounce 
(see Fig. \ref{lepb}) approximately agree or at least are consistent 
with the previous simulations, implying that our code can correctly 
follow the collapse until the bounce.
Also, the radial profiles of the infall velocity, the density, 
and the entropy per baryon just after the bounce show good agreements with 
the previous studies.
No such good agreement was reported in the previous
simulations\cite{Kotake03,PhD} where simple leakage schemes based on
the single neutrino-trapping density were adopted. 
Quantitatively, the negative gradients of the entropy per baryon and the
lepton fraction are little bit steeper in the present simulation than
those in 1D full Boltzmann simulations. 
The reason may be partly because the {\it transfers} of lepton-number 
and energy are not fully solved in the present leakage scheme.
Except for this small quantitative difference, the two results agree well.

For validating a scheme for the neutrino cooling, agreement of
the neutrino luminosities with those by 1D full Boltzmann simulation
should be particularly checked because they depend on both
implementations of weak interactions 
(especially electron capture in the present case) and treatments of
neutrino cooling (the detailed leakage scheme). 
Also, accurate computation of the neutrino luminosities is required
for astrophysical applications, because neutrinos carry away the most 
of energy liberated during the collapse as the main cooling source and
can be primary observable.
Our results, in particular the duration and the peak luminosity of the 
neutrino bursts, agree approximately with those in the previous 
simulations. Again, no such good agreement was reported in 
the previous simulation\cite{Kotake03,PhD}. 

The shock stall-radius is $\approx 80$ km. This value is consistent 
with (although slightly smaller than) that in Liebend\"orfer et al.
\cite{Lieb05b} ($R_{\rm stall} \approx 85$ km) 
and smaller than that in Sumiyoshi et al.\cite{Sumi05} 
($R_{\rm stall}\approx 100$ km).
This is likely because in our leakage scheme, neutrino heating 
is not taken into account.

To summarize, the results in the present simulation agree well with
those in the previous 1D Boltzmann simulations qualitatively.
Quantitatively, the present results agree approximately with those in
the previous 1D Boltzmann simulations.
We can obtain approximately correct results with a not
computationally expensive scheme without solving the Boltzmann equation.
Thus, the present code may be adopted, as a first step, 
to other multidimensional simulations such as the rotating
stellar collapse to a black hole and mergers of compact binaries.

\subsubsection{Comparison of the results after the convection sets in}

In this section, we compare our results in the convective phase 
with those in the two-dimensional (2D) Newtonian 
simulations\cite{BF93,Herant94,BHF95,KJM96,MJ97,Mezza98a,Dessart06,Buras06} 
in which a wide variety of approximations were adopted for the treatment
of neutrinos.

In the present simulation, we have found both the vigorous convective
activities (the proto-neutron star convection) 
and the enhancement of neutrino luminosities 
due to the
convection. These features agree approximately with those in 
the previous 2D simulations with a fluid-like treatment of
neutrinos\cite{Herant94} and with radial ray-by-ray, gray flux-limited
diffusion approximation of neutrino transfers\cite{BF93,BHF95,KJM96}.
In a spherically symmetric, gray flux-limited diffusion
scheme\cite{Mezza98a}, by contrast, only mildly active convection
was found and no enhancement in the neutrino luminosities was observed.

Note that the transport of energy and lepton number by neutrinos can
flatten the negative gradients of entropy and lepton
fraction, and as a result, the convection will be suppressed.
In purely hydrodynamic simulations without neutrino 
processes\cite{MJ97,Mezza98a} (using a postbounce core obtained 
in 1D simulations with neutrinos),
the proto-neutron star convection is strongly activated. 
In the radial ray-by-ray simulations\cite{BF93,BHF95,KJM96}, the
transfer of neutrinos in the angular direction is not taken into account
and the stabilizing effect is underestimated, resulting in the
proto-neutron star convection with the enhancement of neutrino 
luminosities. In the spherically symmetric simulation\cite{Mezza98a}, 
the transfer of neutrino in the angular direction is assumed to occur 
fast enough to make the neutrino distribution function spherically 
symmetric, and consequently, the stabilizing effect is overestimated.

Recently, Buras et al.\cite{Buras06} performed simulations with a
modified ray-by-ray, multi-group scheme in which some part of the
lateral components are included, and found that the proto-neutron star
convection indeed sets in but has minor effects on the enhancement of
the neutrino luminosities. Dessart et al.\cite{Dessart06} performed
simulations employing a 2D multi-group flux-limited diffusion scheme and
found similar results as in Buras et al.
Thus, although the proto-neutron star convection indeed occurs, 
its influence on enhancing the neutrino luminosities may be minor.
The strong convective activities and the enhancement of neutrino luminosities
found in the present simulation should be considered as the maximum ones. 

Note that it is in intermediate regions ($\tau_{\nu} \sim 1$) that the
stabilizing effect due to the neutrino transfer works efficiently:
At higher density region with $\tau_{\nu} \gg 1$, neutrinos cannot efficiently
transport the energy and the lepton number due to the large opacities;
At lower density region with $\tau_{\nu} \ll 1$, on the other hand, neutrinos
carry away the energy and the lepton number without interacting with the
matter. Therefore a careful and detailed treatment of the neutrino
transfer is required to clarify the degree of the stabilizing effect
and the convection, although such a computationally expensive sumulation
is beyond the scope of this paper.

The present result that the proto-neutron star convection occurs 
qualitatively agrees with the recent simulations with detailed neutrino
transfer\cite{Dessart06,Buras06}. If simulations are perfomed keeping 
in mind that the stabilizing effect due to the neutrino transfer is not 
taken into account in the present scheme, the present code will be acceptable
to explore the the rotating stellar collapse to a black hole and mergers of
compact binaries.

\subsection{Gravitational radiation}\label{cGW}

\begin{figure}[t]
  \begin{center}
    \includegraphics[scale=1.0]{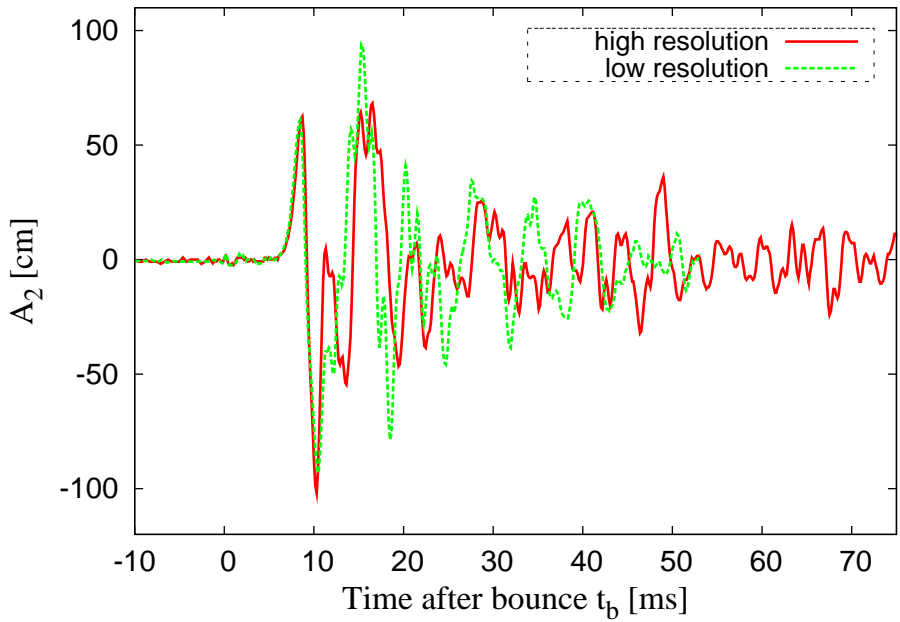}
  \end{center}
  \caption{Gravitational wave quadrupole amplitude $A_{2}$ 
    due to the prompt convection as a
   function of post bounce time $t_{b}$.
    The results for the finer grid resolution (solid curve) 
    and for the coarser grid resolution 
    (the dotted curves) are shown together.
}\label{gw}
\end{figure}

\begin{figure}[t]
  \begin{center}
    \includegraphics[scale=1.0]{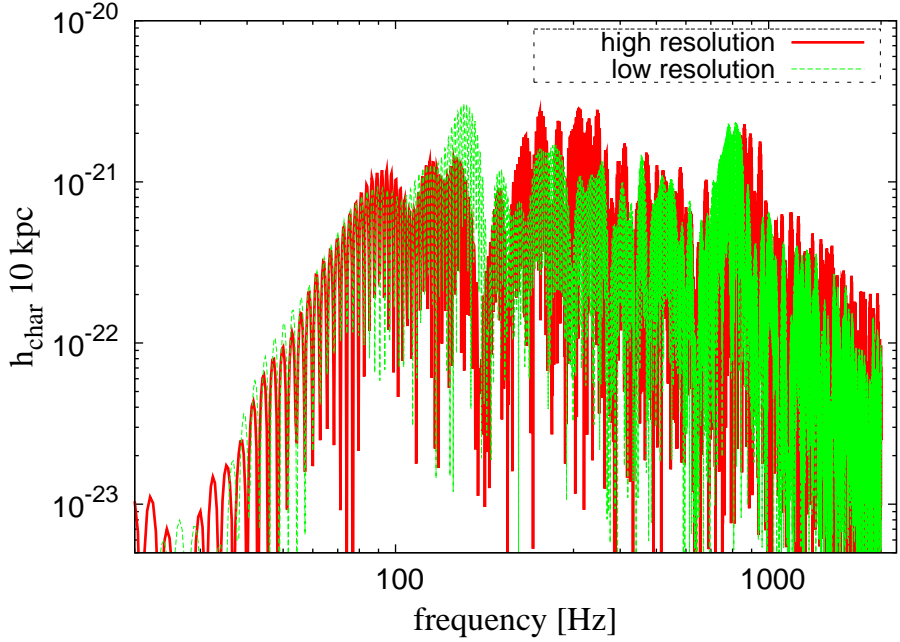}
  \end{center}
  \caption{The frequency spectra of the characteristic gravitational-wave
   strain due to the prompt convection.
    The results for the finer grid resolution (solid curve) 
    and for the coarser grid resolution 
    (the dotted curves) are shown together.
}\label{fgw}
\end{figure}

\begin{figure}[t]
  \begin{center}
    \includegraphics[scale=1.2]{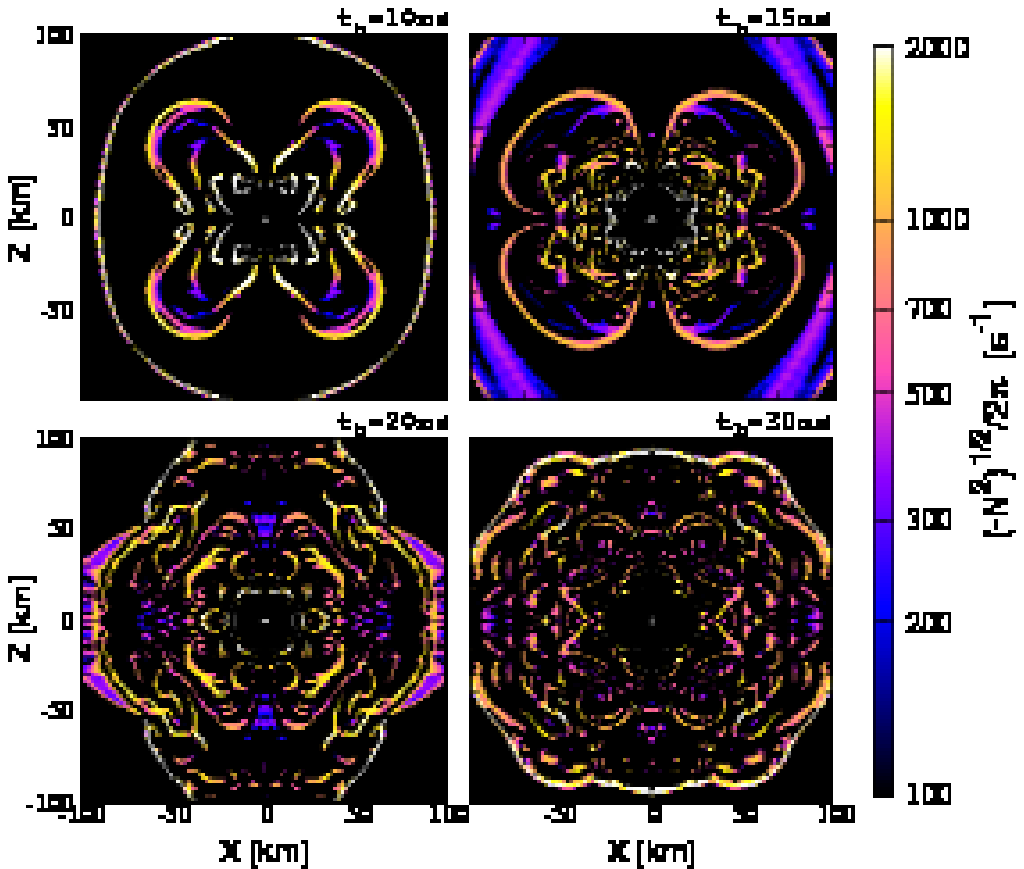}
  \end{center}
  \caption{Snapshots of the contours of $\sqrt{-N^{2}}/2\pi$,  
    in the $x$-$z$ plane at selected time slices.  
}\label{convN}
\end{figure}

Associated with the convective motions, gravitational waves are
emitted. The gravitational waveforms are computed 
using a quadrupole formula\cite{SS03}.
In quadrupole formulae, only the $+$-mode of gravitational waves with 
$l=2$ and $m=0$ is nonzero in axisymmetric spacetime and is written as 
\beq
h_+^{\rm quad} = {\ddot I_{zz}(t_{\rm ret}) - \ddot I_{xx}(t_{\rm ret})
\over r}\sin^2\theta \equiv \frac{A_{2}(t)}{r}\sin^{2}\theta,
\label{quadr}
\eeq
where $I_{ij}$ denotes a quadrupole moment, $\ddot I_{ij}$ 
its second time derivative, and $t_{\rm ret}$ a retarded time. 
In fully general relativistic and dynamical spacetime, 
there is no unique definition for the quadrupole moment and 
nor is for $\ddot I_{ij}$. Following Shibata and Sekiguchi\cite{SS03}, 
we choose the simplest definition of the form
\beq
I_{ij} = \int \rho_* x^i x^j d^3x. 
\eeq
Then, using the continuity equation, 
the first time derivative can be written as 
\beq
\dot I_{ij} = \int \rho_* (v^i x^j +x^i v^j)d^3x,
\eeq
and $\ddot I_{ij}$ is computed by the finite differencing of the numerical result 
for $\dot I_{ij}$.
In the following, we present $A_{2}$, which 
provides the amplitude of a given mode
measured by an observer located in the most optimistic direction (in the
equatorial plane).
We also calculate the characteristic gravitational-wave
strain\cite{FH98}, 
\beq
h_{\rm char}(f) \equiv \sqrt{
\frac{2}{\pi^{2}} \frac{G}{c^{3}}\frac{1}{D^{2}}\frac{dE}{df}},
\eeq \label{hchar}
where
\beq
\frac{dE}{df} = \frac{8\pi^{2}}{15}\frac{c^{3}}{G} f^{2}
\left| \tilde{A}_{2}(f) \right|^{2}
\eeq 
is the energy power spectra of the gravitational radiation
and
\beq
\tilde{A}_{2}(f) = \int A_{2}(t) e^{2\pi i f t} dt.
\eeq

Figure \ref{gw} shows $A_{2}(t)$. Because
the system is initially spherically symmetric,
no gravitational radiation is emitted before the onset of the
convection. 
When the proto-neutron star convection sets in at $\approx 10$ ms 
after the bounce, gravitational waves start to be emitted. 
The peak amplitudes are $A_{2} \sim 100$ cm. 
After the peak is reached, gravitational waves generated by the
smaller-scale convective motions are emitted 
with $A_{2} \approx 50$ cm. 

Figure \ref{fgw} shows the spectra of $h_{\rm char}$ due to the
convective motions. 
In contrast to the spectra due to the core bounce 
(e.g. Refs.~\citen{Dimm02} and \citen{Sekiguchi05}), 
there is no dominant peak frequency in the power spectra. 
Instead, several maxima for the frequency range $100$--$1000$ Hz are
present. Note that for gravitational waves
due to the core bounce, the characteristic peak frequency is associated
with the bounce timescale of the core.
The effective amplitude of gravitational waves observed in the
most optimistic direction is 
$h_{\rm char} \approx$ 6--$8 \times 10^{-21}$ for an
event at a distance of 10 kpc, which is as large as that emitted at the
bounce of rotating core collapse\cite{Dimm02}. 

To check that gravitational waves are indeed originated by the convective
motions, we calculate the frequency $\sqrt{-N^{2}}/2\pi$ (see Eq. (\ref{VB0}))
as shown in Fig. \ref{convN}. This frequency is in good agreement with the
gravitational-wave frequency, implying that gravitational waves 
are indeed due to the convective activities.

M\"uller and Janka\cite{MJ97} investigated gravitational waves
due to the convective motion inside the proto-neutron star. It is
interesting to compare our results with theirs. They adopted a
post-bounce model of Hillebrandt\cite{Hilleb87}. 
They put an inner boundary at radius $r_{\rm in} = 15$ km and assumed the
hydrostatic equilibrium there.
They do not include neutrino transfer while a sophisticated EOS is adopted.
They found qualitatively similar results to ours.
According to their results,
the maximum amplitude of the quadrupole mode is $A_{2} \approx 100$
cm, which agrees well with our results.
The spectrum of the gravitational-wave strain has several maxima for
$f=50$--500 Hz with the maximum value of $h_{\rm char} \approx 3\times
10^{-21}$.
The peaks in $h_{\rm char}$ are distributed for higher
frequency side in our results probably due to the general relativistic
effects. We note that a similar general relativistic effect is observed
for gravitational waves at the bounce phase\cite{Dimm02}.
These facts show that for deriving quantitatively correct spectra of
gravitational waves, fully general relativistic simulations are necessary.

\subsection{Numerical accuracy}

\begin{figure}[t]
  \begin{center}
    \includegraphics[scale=1.1]{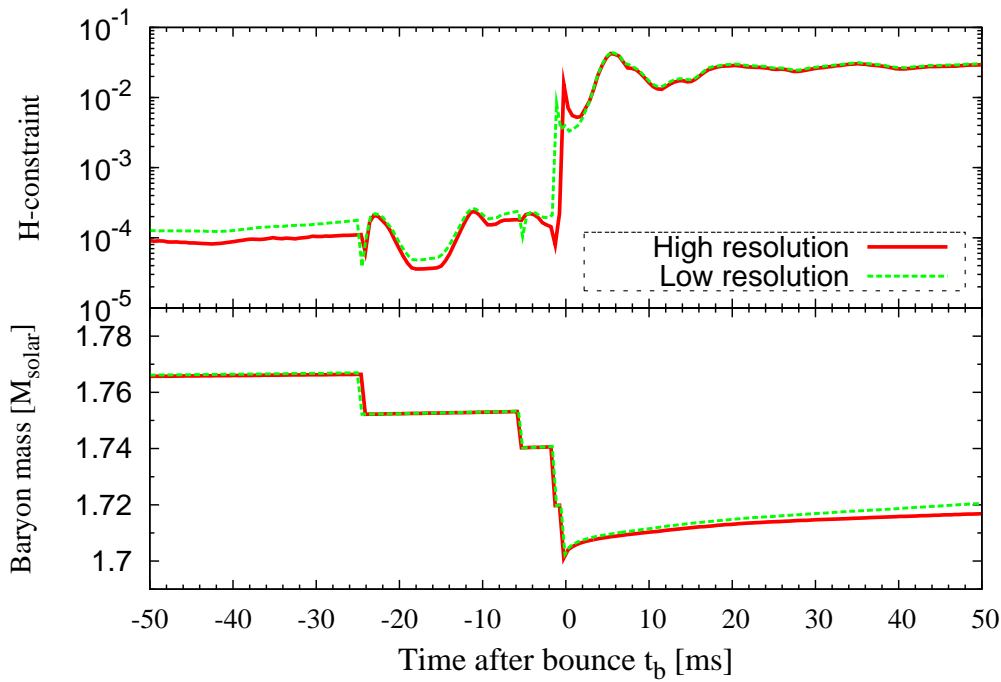}
  \end{center}
  \caption{Evolution of the averaged violation of the Hamiltonian
    constraint (upper panel) and baryon mass conservation (lower panel).}
    \label{ham-mass-b}
\end{figure}

In Figs. \ref{rho-alp}--\ref{nlum} we show the results both in 
the higher resolution (solid curves) and in the lower 
resolution (dashed curves).  
The radial profiles of the two resolutions are almost identical, showing
that convergent results are obtained in the present simulation 
(see Fig. \ref{qalx}).
In the time evolution of neutrino luminosities (see Fig. \ref{nlum}),
the two results are almost identical before the convective activities set in.
In the later phase, on the other hand, the two results show slight
disagreement. Because the convection and the turbulence can occur in an
infinitesimal scale length, the smaller-scale convection and turbulence
are captured in the finer grid resolution. However, the influence of the
grid resolution on the neutrino luminosities is minor because the
convection and turbulence are strongly activated in the region above the 
neutrino sphere (see the contours of the electron fraction and the entropy
in Fig. \ref{con}). On the other hand, most of the neutrinos are 
emitted from the region inside the neutrino sphere
(see the contour of the local neutrino energy emission rate in
Fig. \ref{con}).

The effect of the grid resolution can be seen in gravitational waves. 
In Figs. \ref{gw} and \ref{fgw} we show the quadrupole mode $A_{2}(t)$ and
the characteristic strain $h_{\rm char}(f)$ both in the higher
resolution (solid curves) and in the lower resolution (dashed curves).
After the formation of the lepton-rich, hot finger at $\approx 10$ ms
after the bounce (see \S~\ref{convective_activities}), convective
activities set in. Then, disagreement of $A_{2}(t)$ between the finer
and the coarser grid resolutions becomes noticeable (see Fig. \ref{gw}).
The characteristic peaks of $h_{\rm char}(f)$ in higher frequencies
($f\sim 200$--500 Hz) are more prominent (see Fig. \ref{fgw}). 
It is likely to be because the smaller-scale turbulant motions are captured 
in the finer grid resolution.

To check the accuracy of our numerical results,
the violation of the Hamiltonian constraint are calculated, 
which is written as 
\beq
H=-8 \psi^{-5} \biggl[
\tilde \Delta \psi - {\psi \over 8}\tilde R +2\pi \rho_{h} \psi^5
+{\psi^5 \over 8}\tilde A_{ij}\tilde A^{ij}-{\psi^ 5 \over 12}K^2\biggr],
\eeq 
where $\psi \equiv e^{\phi}$, and
$\tilde \Delta$ denotes the Laplacian with respect to $\tilde \gamma_{ij}$. 
In this paper, the averaged violation is defined according to\cite{Shibata03}
\beq
{\rm ERROR}={1 \over M_*} \int \rho_* |V| d^3x,
\eeq
where $M_{*}$ is the rest-mass density of the core (see Eq. \ref{restM})
\beq
V={\displaystyle \tilde \Delta \psi - {\psi \over 8}\tilde R +2\pi E \psi^5
+{\psi^5 \over 8}\tilde A_{ij}\tilde A^{ij}-{\psi^ 5 \over 12}K^2
\over \displaystyle
|\tilde \Delta \psi| + \Big|{\psi \over 8}\tilde R \Big| +2\pi \rho_{h} \psi^5
+{\psi^5 \over 8}\tilde A_{ij}\tilde A^{ij}+{\psi^ 5 \over 12}K^2}. 
\eeq

Namely, we use $\rho_*$ as a weight factor for the average.
This weight factor is introduced to monitor whether the main bodies of
the system (proto-neutron stars and inner cores), 
in which we are interested, are accurately computed or not.

We display the time evolution of the Hamiltonian-constraint violation 
and the conservation of the baryon mass of the system in Fig. \ref{ham-mass-b}.  
Several discontinuous changes in the Hamiltonian-constraint violation 
and the conservation of the baryon mass originate from the regridding 
procedures in which some matters of the outer region are discarded. 

Before the bounce, the baryon mass is well conserved and the 
Hamiltonian-constraint violation is very small as $\sim 10^{-4}$. 
After the bounce, the violation of the baryon-mass-conservation and the
Hamiltonian constraint is enhanced due to the existence of shock waves
where the hydrodynamic scheme becomes essentially a first-order scheme.
The convergence of the baryon-mass-conservation and
the Hamiltonian-constraint violation also becomes worse in 
the convective phase.
However, the degree of violation of the Hamiltonian constraint and the
baryon-mass-conservation is small and we may believe that the numerical
results obtained in the paper are reliable.

\section{Summary and Discussion}
\label{S_Summary}

\subsection{Summary}

In this paper, we present a fully general relativistic hydrodynamic code
in which a finite-temperature EOS and neutrino cooling are implemented
for the first time.
Because the characteristic timescale of weak interaction processes
$t_{\rm wp} \sim \vert Y_{e}/\dot{Y}_{e} \vert$ (WP timescale) is much 
shorter than the dynamical timescale $t_{\rm dyn}$ in hot dense matters, 
stiff source terms appear in the equations.
In general, an implicit scheme may be required to solve them \cite{Bruenn85}.
However, it is not clear whether implicit schemes do work or not in the 
relativistic framework.
The Lorentz factor is coupled with the rest-mass density and the energy 
density. The specific enthalpy is also coupled with the momentum. 
Due to these couplings, it is not straightforward to recover the
primitive variables and the Lorentz factor from conserved quantities. 
Taking account of these facts, we proposed an explicit method to solve
all the equations noting that the characteristic timescale of neutrino
leakage from the system $t_{\rm leak}$  is much longer than $t_{\rm wp}$
and is comparable to $t_{\rm dyn}$.

By decomposing the energy-momentum tensor of neutrinos into the
trapped-neutrino and the streaming-neutrino parts, the hydrodynamic 
equations can be rewritten so that the source terms are characterized by
the leakage timescale $t_{\rm leak}$ (see Eqs. (\ref{T_Eq_M}) and (\ref{T_Eq_nuS})).
The lepton-number conservation equations, on the other hand, include the
source terms characterized by the WP timescale. 
Taking account of these facts, {\it limiters} for the stiff source
terms are introduced to solve the lepton-number conservation equations
explicitly (see \S~\ref{Lepton}).
In the numerical relativistic hydrodynamics, it is required to
calculate the primitive variables and the Lorentz factor from the 
conserved quantities. In this paper, we develop a robust and stable 
procedure for it (\S~\ref{Reconst}).

To check the validity of the numerical code, we performed a simulation
of spherical stellar core collapse. As initial conditions, we adopted 
the $15M_{\odot}$ spherical model with the solar metallicity
computed by Woosley et al.\cite{WHW02}
After the shock formation and propagation, the shock wave suffers from 
the severe reduction of its energy due to neutrino burst emission when 
the shock wave passes the neutrino-sphere. Eventually, the shock wave 
stalls soon after it passes through the neutrino sphere. 
The neutrino burst makes negative gradients of the entropy and $Y_{l}$ 
above the neutrino sphere. Because such configuration is
convectively unstable, vigorous convective motions are induced.
All these properties agree qualitatively with those by the resent 2D
Newtonian simulations\cite{Dessart06,Buras06}.

We also compared our results with those in the previous simulations.
Before the convection sets in, we compare our result with those in 
the state-of-the-art 1D Boltzmann simulations in full general
relativity\cite{Lieben01,Lieb04,Lieb05,Sumi05}.
As shown in this paper, the radial structure of the core and the
neutrino luminosities agree qualitatively well with those in 
their simulations. Quantitatively, they also agree approximately with
the previous results. 

After the convection sets in, we compare our result with those in 
2D Newtonian
simulations\cite{BF93,Herant94,BHF95,KJM96,MJ97,Mezza98a,Dessart06,Buras06}.
Our result that the proto-neutron star convection occurs
agree qualitatively with that in the previous
simulations\cite{BF93,Herant94,BHF95,KJM96,MJ97,Dessart06,Buras06}.
However, quantitative properties show disagreement because the transfer
of neutrinos are not fully solved in the present scheme.
Note that the transport of energy and lepton-number by neutrinos can
flatten the negative gradients of entropy and lepton fraction,
stabilizing the convection.
Therefore the convective activities obtained in the present simulation
should be considered as the maximum ones.

If we keep in mind the above facts and note the good agreements of the
radial structure and neutrino luminosities, the present implementation
will be applied to simulations of rotating core collapse to a black hole
and mergers of binary neutron stars as a first step towards more
sophisticated models. A detailed treatment of the neutrino transfer 
is required to determine the degree of stabilizing effect, but this is
far beyond the scope of this paper.

Gravitational waves emitted by the convective motions are also
calculated. The gravitational-wave amplitude is 
$\approx 3 \times 10^{-21}$ for an event of the distance $10$ kpc.
Reflecting the contributions of multi-scale eddies with characteristic
overturn timescale 1--10 ms, the energy power spectrum shows several
maxima distributed in $f\approx 100$--1000 Hz. We compare our results 
with those in M\"uller and Janka\cite{MJ97} in which a similar 
calculation (but in Newtonian gravity) is performed. 
The maximum amplitude of gravitational waves 
in our results agrees well with that in M\"uller and Janka.
The several maxima in the energy power spectrum are distributed at 
higher-frequency side in our results due to the general relativistic
effects, showing that fully general relativistic simulations are
necessary for the accurate calculation of gravitational-wave spectra.

\subsection{Discussions}

Because the present implementation of the microphysics is simple and
explicit, it has advantage that the individual microphysical processes
can be easily improved and sophisticated.
For example, the neutrino emission via the electron capture can be
easily sophisticated as follows.
To precisely calculate the electron capture rate, the complete
information of the parent and daughter nuclei is required.
In EOSs currently available, however, a representative single-nucleus 
average for the true ensemble of heavy nuclei is adopted. 
The representative is usually the most abundant nucleus.
The problem in evaluating the capture rate is that the nuclei
which cause the largest changes in $Y_{e}$ are neither the most
abundant nuclei nor the nuclei with the largest rates, but the
combination of the two. In fact, the most abundant nuclei tend to have
small rates because they are more stable than others, and the fraction 
of the most reactive nuclei tend to be small\cite{AFWH94,Janka07a}.
Assuming that the nuclear statistical equilibrium (NSE) is achieved, the
electron capture rates under the NSE ensemble of heavy nuclei may be 
calculated for a given set of ($\rho$, $Y_{e}$, $T$).
Such a numerical rate table can be easily employed in the present
implementation.

Also, the neutrino cross sections can be improved. As summarized in
Ref.~\citen{Horowitz}, there are a lot of corrections to the neutrino 
opacities. Note that small changes in the opacities may result in much 
larger changes in the neutrino luminosities, because the neutrino energy 
emission rates depend strongly on the temperature, and the temperature 
at the last scattering surface ($\tau_{\nu}\sim \sigma T^{2} \sim 1$) 
changes as $T \sim \sigma^{-1/2}$.
Although the correction terms are in general very complicated, it is 
straightforward to include the corrections into our code. 
Note that the corrections become more important for
higher neutrino energies. Therefore, the correction terms might play
a crucial role in the collapse of population III stellar core and the
formation of a black hole, in which very high temperatures ($T>100$ MeV)
will be achieved. A study to explore the importance of these corrections in 
the case of black hole formation is ongoing.

As briefly described in the introduction, one of the main drawbacks 
in the present implementation of the neutrino cooling is that the 
{\it transfer} of neutrinos are not solved. Although {\it fully}
solving the transfer equations of neutrinos is far beyond the scope of 
this paper, there are a lot of rooms for improvements in the treatment 
of the neutrino cooling. For example, the relativistic moment 
formalism\cite{AS72,Thorne81}, in particular the so-called M1 closure
formalism, may be adopted.
For this purpose, a more sophisticated treatment of the closure relation
for $P_{\alpha \beta}$ is required. 
We plan to implement a relativistic M1 closure formalism for the
neutrino transfer in the near future.

To conclude, the present implementation of microphysics in fully general
relativistic, multidimensional code works well and has a wide variety of
applications. We are now in the standpoint where simulations
of stellar core collapse to a black hole and merger of compact stellar binaries 
can be performed including microphysical processes. 
Fruitful scientific results will be reported in the near future.

\section*{acknowledgments}

The author thanks M. Shibata for valuable discussions and careful
reading of the manuscript, 
and L. Rezzolla, and K. Sumiyoshi for valuable discussions.
He also thanks T. Shiromizu and T. Fukushige for their grateful aids.
He thanks S. E. Woosley,  A. Heger and T. A. Weaver for providing the
presupernova core used in the present simulation as an initial condition.
Numerical computations were performed on the NEC SX-9 at the data
analysis center of NAOJ and on the NEC SX-8 at YITP in Kyoto University. 
This work is partly
supported by the Grant-in-Aid of the Japanese Ministry of Education, Science,
Culture, and Sport (21018008,21105511).

\appendix

\section{Electron and positron captures}\label{A_EPC} \label{EleCap}

In this section, we briefly summarize our treatments of electron and positron
captures which are based on Ref.~\citen{FFN85} and give the explicit
forms of $\gamma^{\rm ec}_{\nu e}$, $\gamma^{\rm pc}_{\bar{\nu} e}$,
$Q^{\rm ec}_{\nu e}$, and $Q^{\rm pc}_{\bar{\nu} e}$ appeared in 
Eqs. (\ref{gnlocal}), (\ref{galocal}), and (\ref{Qlocal}),
for the purpose of convenience.

\subsection{The electron and positron capture rates 
$\gamma^{\rm ec}_{\nu e}$ and $\gamma^{\rm pc}_{\bar{\nu} e}$}
The 'net' electron fraction is written as $Y_{e} = Y_{-} - Y_{+}$ 
where $Y_{-}$ ($Y_{+}$) denotes the number of electrons (positrons) 
per baryon including pair electrons. 
Then the electron-neutrino number emission rate by the electron capture
and the electron-anti-neutrino number emission rate by the positron capture
are given by 
\beqn
&& \gamma^{\rm local}_{\nu e} = -\dot{Y}_{-} = 
  -(\dot{Y}_{-}^{f} +\dot{Y}_{-}^{h}), \\
&& \gamma^{\rm local}_{\bar{\nu} e} = -\dot{Y}_{+} = 
  -(\dot{Y}_{+}^{f} +\dot{Y}_{+}^{h}),
\eeqn
where the electron and positron capture rates are decomposed into two
parts, capture on by free nucleons (with superscript $f$) and on heavy
nuclei (with superscript $h$). 
In the following, we present the explicit forms of 
$\dot{Y}_{-}^{f}$, $\dot{Y}_{+}^{f}$, $\dot{Y}_{-}^{h}$, and
$\dot{Y}_{+}^{h}$.

\subsection{Capture on free nucleons $\dot{Y}^{f}$}

The electron capture rate (including the contribution of the inverse
reaction of the neutrino capture) on free nucleons ($\dot{Y}_{-}^{f}$) is
given by 
\beq
\dot{Y}_{-}^{f} = X_{n}\lambda^{\nu_{e} {\rm c},f} 
-X_{p}\lambda^{{\rm ec},f},
\label{dot_Yef}
\eeq
where $\lambda^{{\rm ec},f}$ is the specific 
electron capture rate on free protons, 
$\lambda^{\nu_{e} {\rm c},f}$ is the specific electron-neutrino capture
rate on free neutrons, 
and  $X_{p}$ and $X_{n}$ are the mass fraction of free proton and neutron,
respectively. 
Based on a balance argument\cite{FFN85}, 
one can show that $\lambda^{\nu_{e} {\rm c},f}$
is related to $\lambda^{{\rm ec},f}$ by 
\beq 
\lambda^{\nu_{e} {\rm c},f} = 
\exp \left(
\eta_{\nu e} -\eta_{e} - \frac{\delta m}{k_{B}T}
\right)
\lambda^{{\rm ec},f},
\label{ecnc}
\eeq
where $\eta_{\nu e}$ and $\eta_{e}$ are the chemical potentials of
electron neutrinos and electrons in units of $k_{B}T$ and 
$\delta m = (m_{n} - m_{p})c^{2}$.
Furthermore, we use the following relation for non-degenerate free nucleons,
\beq
X_{n} \approx X_{p} 
\exp \left( \eta_{n}-\eta_{p}+\frac{\delta m}{k_{B}T} \right), 
\eeq
where $\eta_{n}$ and $\eta_{p}$ are the chemical potentials of
free neutrons and free protons in units of $k_{B}T$.
Then we obtain
\beq
\dot{Y}_{-}^{f} = \left[
\exp \left(\eta_{\nu e}-\eta_{e} + \eta_{n}-\eta_{p} \right)
-1 \right] X_{p}\lambda^{{\rm ec},f}.
\eeq

The positron capture rate 
(including the contribution of the inverse reaction) on free nucleons is
similarly given by   
\beq
\dot{Y}_{+}^{f} = X_{p}\lambda^{\bar{\nu}_{e} {\rm c},f} -X_{n}\lambda^{{\rm pc},f}
= 
\left[ \exp 
\left(\eta_{\bar{\nu}e} +\eta_{e} + \eta_{p}-\eta_{n} \right)
- 1 \right] X_{n} \lambda^{{\rm pc},f}, \label{dot_Ypf}
\eeq
where $\eta_{\bar{\nu} e}$ is the chemical potential of
electron-anti-neutrinos in units of $k_{B}T$, 
$\lambda^{\rm pc}$ is the specific positron capture rate on free
neutrons, and $\lambda^{\bar{\nu}_{e} {\rm c},f}$ is the specific
electron-anti-neutrino capture rate on free protons.

\subsection{Capture on heavy nuclei $\dot{Y}^{h}$}

The electron capture rate (including the contribution of the inverse
reaction of the neutrino capture) on a heavy nucleus of mass number 
$A$ ($\dot{Y}_{-}^{h}$) is given by\cite{FFN85}
\beq
\dot{Y}_{-}^{h} = \frac{X_{D}}{A}\lambda^{\nu_{e} {\rm c}, h} 
                 -\frac{X_{P}}{A}\lambda^{{\rm ec}, h}, 
\label{dot_Yeh}
\eeq
where $\lambda^{{\rm ec},h}$ is the specific electron capture rate on the parent
nucleus (mass fraction $X_{P}$),
$\lambda^{\nu_{e} {\rm c},h}$ is the specific electron neutrino capture rate on the
daughter nucleus (mass fraction $X_{D}$),
and $A$ is the atomic mass of the parent and daughter nuclei.
In the present simulations, we set $X_{D} = X_{P} = X_{A}$.
Then, under the assumption of nuclear statistical equilibrium,
one may approximate the capture rate on heavy nuclei as\cite{FFN85},
\beq
\dot{Y}_{-}^{h} 
\approx
\left[  
\exp \left(
\eta_{\nu e} - \eta_{e} + \eta_{n}-\eta_{p}
\right) - 1 
\right] \frac{X_{A}}{A} \lambda^{{\rm ec},h}. 
\label{dot_Yeh3}
\eeq

Similarly, the positron capture rate (including the contribution of the
inverse reaction) on heavy nuclei ($\dot{Y}_{+}^{h}$) is given by 
\beq
\dot{Y}_{+}^{h} = \frac{X_{D}}{A}\lambda^{\bar{\nu}_{e} {\rm c}, h} 
                 -\frac{X_{P}}{A}\lambda^{{\rm pc}, h}
\approx
\left[  \exp \left(
\eta_{\bar{\nu}e} + \eta_{e} + \eta_{p}-\eta_{n}
\right) - 1
\right]\frac{X_{A}}{A}\lambda^{{\rm pc},h} .
\label{dot_Yph}
\eeq

\subsection{The specific capture rate $\lambda$}

The specific electron and positron capture rates on 
free nucleons and on heavy nuclei and are written 
in the same form as\cite{FFN85}
\beqn
&& 
\lambda^{{\rm ec}, f} = 
\frac{\ln 2}{\langle ft \rangle _{\rm eff}^{{\rm ec},f}} I^{{\rm ec},f},
\ \ \ \ \ 
\lambda^{{\rm pc}, f} = 
\frac{\ln 2}{\langle ft \rangle _{\rm eff}^{{\rm pc},f}} I^{{\rm pc},f}, 
\\ 
&&
\lambda^{{\rm ec}, h} = 
\frac{\ln 2}{\langle ft \rangle _{\rm eff}^{{\rm ec},h}} I^{{\rm ec},h},
\ \ \ \ \ 
\lambda^{{\rm pc}, h} = 
\frac{\ln 2}{\langle ft \rangle _{\rm eff}^{{\rm pc},h}} I^{{\rm pc},h}, 
\eeqn
where $I^{{\rm ec},f}$ and $I^{{\rm pc},f}$ are the phase space factors
for the electron and positron captures on free electrons,
and $I^{{\rm ec},h}$ and $I^{{\rm pc},h}$ are those on heavy nuclei.
$\langle ft \rangle _{\rm eff}$'s are the effective
$ft$-values introduced by Fuller et al.\cite{FFN85}, 
which is essentially the same as the square of 
the nuclear transition matrix.

The phase space factors are given by
\beqn
\!\!\!\!\!\!\!\!\!
I^{{\rm ec},f} \!&=&\!
\left( \frac{k_{B}T}{m_{e}c^{2}} \right)^{5}
\int_{\eta_{0}}^{\infty} 
\eta^{2}(\eta + \zeta^{{\rm ec},f})^{2}
\frac{1}{1+e^{\eta - \eta_{e}}}
\left[
1- \frac{1}{1+e^{\eta -\eta_{\nu e} +\zeta^{{\rm ec},f} }}
\right]
d\eta, \label{def_Ie}
\\
\!\!\!\!\!\!\!\!\!
I^{{\rm pc},f} \!&=&\! 
\left( \frac{k_{B}T}{m_{e}c^{2}} \right)^{5}
\int_{\eta_{0}}^{\infty} 
\eta^{2}
(\eta + \zeta^{{\rm pc},f})^{2}
\frac{1}{1+e^{\eta + \eta_{e}}}
\left[
1- \frac{1}{1+e^{\eta -\eta_{\bar{\nu}e} +\zeta^{{\rm pc},f} }}
\right]
d\eta, \label{def_Ip}
\\
\!\!\!\!\!\!\!\!\!
I^{{\rm ec},h} \!&=&\!
\left( \frac{k_{B}T}{m_{e}c^{2}} \right)^{5}
\int_{\eta_{0}}^{\infty} 
\eta^{2}(\eta + \zeta^{{\rm ec},h})^{2}
\frac{1}{1+e^{\eta - \eta_{e}}}
\left[
1- \frac{1}{1+e^{\eta -\eta_{\nu e} +\zeta^{{\rm ec},h} }}
\right]
d\eta, \label{def_Ieh}
\\
\!\!\!\!\!\!\!\!\!
I^{{\rm pc},h} \!&=&\! 
\left( \frac{k_{B}T}{m_{e}c^{2}} \right)^{5}
\int_{\eta_{0}}^{\infty} 
\eta^{2}
(\eta + \zeta^{{\rm pc},h})^{2}
\frac{1}{1+e^{\eta + \eta_{e}}}
\left[
1- \frac{1}{1+e^{\eta -\eta_{\bar{\nu}e} +\zeta^{{\rm pc},h} }}
\right]
d\eta, \label{def_Iph}
\eeqn 
where $\zeta^{{\rm ec},f}$, $\zeta^{{\rm pc},f}$, 
$\zeta^{{\rm ec},h}$, and $\zeta^{{\rm pc},h}$ are the nuclear mass-energy
differences for electron capture and positron capture in units of $k_{B}T$. 
The nuclear mass-energy differences for capture on free nuclei are given by
\beq
\zeta^{{\rm ec},f} 
= -\zeta_{n}^{{\rm pc},f}
\approx 
\eta_{p} - \eta_{n}.
\eeq
We follow Fuller et al.\cite{FFN85} for the nuclear
mass-energy differences for capture on heavy nuclei:
In the case of $N<40$ or $Z>20$ (referred to as 'unblocked' case),
we set
\beq
\zeta^{{\rm ec},h} 
= -\zeta^{{\rm pc},h}
\approx 
\eta_{p} - \eta_{n}.
\eeq
In the case of $N\ge40$ or $Z\le20$ (referred to as 'blocked' case),
on the other hand, we set
\beqn
\zeta^{{\rm ec},h} 
&\approx&
\eta_{p}-\eta_{n} - \frac{5 ({\rm MeV})}{k_{B}T}, \\
\zeta^{{\rm pc},h} 
&\approx& 
-\eta_{p}+\eta_{n} + \frac{5 ({\rm MeV})}{k_{B}T}.
\eeqn
Then, the threshold value of the electron and positron captures is given by
$\eta_{0} = m_{e}c^{2}/(k_{B}T)$ for $\zeta > -m_{e}c^{2}/(k_{B}T)$
and $\eta_{0} = |\zeta|$ for $\zeta < -m_{e}c^{2}/(k_{B}T)$ where
we have dropped the superscripts 'ec', 'pc', '$f$', and '$h$' in $\zeta$
for simplicity.

The effective $ft$-value of electron or positron capture on free nuclei
is given by (e.g. Ref.~\citen{FFN85}
\beq
\log_{10} \langle ft \rangle_{\rm eff}^{{\rm ec},f} =
\log_{10} \langle ft \rangle_{\rm eff}^{{\rm pc},f} \approx 3.035.
\eeq
We follow Fuller et al.\cite{FFN85} for the effective $ft$-value of
capture on heavy nuclei, who proposed to use 
\beqn
&&
\log_{10} \langle ft \rangle_{\rm eff}^{{\rm ec},h} \approx \left\{
\begin{array}{ccc}
3.2 & {\rm unblocked} & \eta_{e}<|\zeta^{{\rm ec},h}|  \\
2.6 & {\rm unblocked} & \eta_{e}>|\zeta^{{\rm ec},h}|  \\
2.6 + \frac{25.9}{T_{9}} & {\rm blocked} &
\end{array}
\right. , \label{eft1} \\
&&
\log_{10} \langle ft \rangle_{\rm eff}^{{\rm pc},h} \approx \left\{
\begin{array}{ccc}
3.2 & {\rm unblocked} & \eta_{e}<|\zeta^{{\rm pc},h}|  \\
2.6 & {\rm unblocked} & \eta_{e}>|\zeta^{{\rm pc},h}|  \\
2.6 + \frac{25.9}{T_{9}} & {\rm blocked} &
\end{array}
\right. , \label{eft2}
\eeqn
where $T_{9} = T/(10^{9}K)$. 
In this expression, the thermal unblocking effect\cite{CW84}
is readily taken into account.
In the thermal unblocking, it costs $\approx 5.13$ MeV to pull a neutron
out of a filled orbital 1$f_{5/2}$ and place it in the
$gd$-shell\cite{FFN85}. 

\subsection{Energy emission rates 
$Q^{\rm ec}_{\nu e}$ and $Q^{\rm pc}_{\bar{\nu} e}$} 

The neutrino energy emission rates associated with 
electron and positron captures in units of $m_{e}c^{2}$ s$^{-1}$ 
are given by\cite{FFN85}
\beq
\pi^{\rm ec} = \ln 2 \frac{J^{\rm ec}}
{\langle ft \rangle_{\rm eff}^{\rm ec}}, \ \ \ \ \ \ 
\pi^{\rm pc} = \ln 2 \frac{J^{\rm pc}}
{\langle ft \rangle_{\rm eff}^{\rm pc}},
\label{pi_epc}
\eeq
where the phase space factors are given by
\beqn
J^{\rm ec} &=& \left(\frac{k_{B}T}{m_{e}c^{2}}\right)^{6}
\int_{\eta_{0}}^{\infty}
\eta^{2}(\eta + \zeta^{\rm ec})^{3}
\frac{1}{1+e^{\eta - \eta_{e}}}
\left[ 1- 
\frac{1}{1-e^{\eta -\eta_{\nu e} + \zeta^{\rm ec}}}\right]d\eta 
\label{J_ec},\\
J^{\rm pc} &=& \left(\frac{k_{B}T}{m_{e}c^{2}}\right)^{6}
\int_{\eta_{0}}^{\infty}
\eta^{2}(\eta + \zeta^{\rm pc})^{3}
\frac{1}{1+e^{\eta + \eta_{e}}}
\left[ 1- 
\frac{1}{1-e^{\eta -\eta_{\bar{\nu}e} + \zeta^{\rm pc} }}\right]d\eta .
\label{J_pc}
\eeqn
In Eqs. (\ref{pi_epc})--(\ref{J_pc}), we have dropped the superscripts
'$f$' and '$h$' in $\pi^{\rm ec}$, $\pi^{\rm pc}$, $J^{\rm ec}$, 
$J^{\rm pc}$, $\langle ft \rangle_{\rm eff}^{\rm ec}$, 
$\langle ft \rangle_{\rm eff}^{\rm pc}$, $\zeta^{\rm ec}$, and
$\zeta^{\rm pc}$ for simplicity.

The average energy of electron neutrinos produced by electron 
and positron captures is defined, in units of $m_{e}c^{2}$, as
\beq
\langle \epsilon_{\nu e} \rangle^{\rm ec} = 
\frac{J^{\rm ec}}{I^{\rm ec}}, \ \ \ \ \ \ 
\langle \epsilon_{\bar{\nu}e} \rangle^{\rm pc} = 
\frac{J^{\rm ec}}{I^{\rm pc}}.
\eeq
Then, the local neutrino energy emission rates by the electron and
positron captures per unit volume is given by 
\beqn
Q_{\nu e}^{\rm ec} &=& \frac{\rho}{m_{u}}
\left[\,
  X_{p} \langle \epsilon_{\nu e} \rangle^{{\rm ec},f} 
  \lambda^{{\rm ec}, f}
+ \frac{X_{A}}{A} \langle \epsilon_{\nu e} \rangle^{{\rm ec},h}
  \lambda^{{\rm ec}, h}
\right], \\
Q_{\bar{\nu} e}^{\rm pc} &=& \frac{\rho}{m_{u}}
\left[\,
 X_{n} \langle \epsilon_{\bar{\nu}e} \rangle^{{\rm pc},f} 
  \lambda^{{\rm pc}, f}
+\frac{X_{A}}{A} \langle \epsilon_{\bar{\nu}e} \rangle^{{\rm pc},h}
 \lambda^{{\rm pc}, h}\, 
\right].
\eeqn

\section{Neutrino pair processes} \label{A_nupair}

In this section, we briefly summarize our treatment of pair processes of
neutrino emission and give the explicit forms of 
$\gamma^{\rm pair}_{\nu_{e}\bar{\nu}_e}$, 
$\gamma^{\rm plas}_{\nu_{e}\bar{\nu}_e}$, 
$\gamma^{\rm Brems}_{\nu_{e}\bar{\nu}_e}$, 
$\gamma^{\rm pair}_{\nu_{x}\bar{\nu}_x}$, 
$\gamma^{\rm plas}_{\nu_{x}\bar{\nu}_x}$, 
$\gamma^{\rm Brems}_{\nu_{x}\bar{\nu}_x}$, 
$Q^{\rm pair}_{\nu_{e}\bar{\nu}_e}$, 
$Q^{\rm plas}_{\nu_{e}\bar{\nu}_e}$, 
$Q^{\rm Brems}_{\nu_{e}\bar{\nu}_e}$, 
$Q^{\rm pair}_{\nu_{x}\bar{\nu}_x}$, 
$Q^{\rm plas}_{\nu_{x}\bar{\nu}_x}$, and
$Q^{\rm Brems}_{\nu_{x}\bar{\nu}_x}$
appeared in
Eqs. (\ref{gnlocal}), (\ref{galocal}), (\ref{gxlocal}) and
(\ref{Qlocal}), for the purpose of convenience.

\subsection{Electron-positron pair annihilation}

We follow Cooperstein et al. \cite{CHB86} for the rate of pair creation
of neutrinos by the electron-positron pair annihilation.
The number emission rate of $\nu_{e}$ or $\bar{\nu}_{e}$ by the
electron-positron pair annihilation can be written as 
\beq
\gamma_{\nu_{e} \bar{\nu}_{e}}^{\rm pair} =
\frac{m_{u}}{\rho}
\frac{C^{\rm pair}_{\nu_{e}\bar{\nu}_e}}{36\pi^{4}} 
\frac{\sigma_{0}c  }{m_{e}^{2}c^{4}}
\frac{(k_{B}T)^{8} }{(\hbar c)^{6}}
F_{3}(\eta_{e})F_{3}(-\eta_{e})
\langle {\rm block} \rangle_{\nu_{e}\bar{\nu}_{e}}^{\rm pair},
\eeq
where $\sigma_{0} \approx 1.705 \times 10^{-44}$cm$^{-2}$ and 
$C^{\rm pair}_{\nu_{e}\bar{\nu}_{e}} = (C_{V}-C_{A})^{2}+(C_{V}+C_{A})^{2}$ 
with $C_{V} = \frac{1}{2} + 2\sin^{2}\theta_{W}$ and $C_{A} = \frac{1}{2}$.
The Weinberg angle is given by $\sin^{2}\theta_{W} \approx 0.23$. 
Using the average energy of neutrinos produced by the pair annihilation,
\beq
\langle \epsilon_{\nu_{e}\bar{\nu}_{e}} \rangle^{\rm pair} =
\frac{k_{B}T}{2}\left(
\frac{F_{4}(\eta_{e})}{F_{3}(\eta_{e})} +
\frac{F_{4}(-\eta_{e})}{F_{3}(-\eta_{e})}
\right),
\eeq
the blocking factor $\langle {\rm block}
\rangle_{\nu_{e}\bar{\nu}_{e}}^{\rm pair}$ is 
evaluated as
\beq
\langle {\rm block} \rangle_{\nu e}^{\rm pair} \approx
\left[
1+\exp \left(
\eta_{\nu e} - 
\frac{\langle \epsilon_{\nu_{e}\bar{\nu}_{e}} \rangle^{\rm pair}}
{k_{B}T}
\right)
\right]^{-1} 
\left[
1+\exp \left(
\eta_{\bar{\nu} e} - 
\frac{\langle \epsilon_{\nu_{e}\bar{\nu}_{e}} \rangle^{\rm pair}}
{k_{B}T}
\right)
\right]^{-1} .
\eeq
The associated neutrino energy emission rate by the pair annihilation is
given by 
\beq
Q_{\nu_{e}\bar{\nu}_{e}}^{\rm pair} =
\frac{\rho}{m_{u}}\gamma_{\nu_{e}\bar{\nu}_{e}}^{\rm pair}
\langle \epsilon_{\nu_{e}\bar{\nu}_{e}} \rangle^{\rm pair}.
\eeq

Similarly, the number emission rate of $\nu_{x}$ or $\bar{\nu}_{x}$ by the
electron-positron pair annihilation and the associated energy emission
rate are given by
\beqn
&&
\gamma_{\nu_{x} \bar{\nu}_{x}}^{\rm pair} =
\frac{m_{u}}{\rho}
\frac{C^{\rm pair}_{\nu_{x}\bar{\nu}_{x}}}{36\pi^{4}} 
\frac{\sigma_{0}c  }{m_{e}^{2}c^{4}}
\frac{(k_{B}T)^{8} }{(\hbar c)^{6}}
F_{3}(\eta_{e})F_{3}(-\eta_{e})
\langle {\rm block} \rangle_{\nu_{x}\bar{\nu}_{x}}^{\rm pair}, \\
&&
Q_{\nu_{x}\bar{\nu}_{x}}^{\rm pair} = 
\frac{\rho}{m_{u}}\gamma_{\nu_{x}\bar{\nu}_{x}}^{\rm pair}
\langle \epsilon_{\nu_{x}\bar{\nu}_{x}} \rangle^{\rm pair},
\eeqn
where $C_{\nu_{x}\bar{\nu}_{x}} = (C_{V}-C_{A})^{2}+(C_{V}+C_{A}-2)^{2}$.
The average neutrino energy and the blocking factor are given by
\beq
\langle \epsilon_{\nu_{x}\bar{\nu}_{x}} \rangle^{\rm pair} =
\langle \epsilon_{\nu_{e}\bar{\nu}_{e}} \rangle^{\rm pair}
\eeq
and 
\beq
\langle {\rm block} \rangle_{\nu e}^{\rm pair} \approx
\left[
1+\exp \left(
\eta_{\nu x} - 
\frac{\langle \epsilon_{\nu_{x}\bar{\nu}_{x}} \rangle^{\rm pair}}
{k_{B}T}
\right)
\right]^{-1} 
\left[
1+\exp \left(
\eta_{\bar{\nu} x} - 
\frac{\langle \epsilon_{\nu_{x}\bar{\nu}_{x}} \rangle^{\rm pair}}
{k_{B}T}
\right)
\right]^{-1},
\eeq
where note that $\eta_{\bar{\nu} x} = \eta_{\nu x}$.

\subsection{Plasmon decay}

We follow Ruffert et al. \cite{RJS96} for the rate of pair creation of neutrinos
by the decay of transversal plasmons.
The number emission rate of $\nu_{e}$ or $\bar{\nu}_{e}$ can be written as
\beq
\gamma_{\nu_{e}\bar{\nu}_{e}}^{\rm plas} = 
\frac{m_{u}}{\rho}
\frac{C_{V}^{2}}{192 \pi^{3} \alpha_{\rm fine}}
\frac{\sigma_{0}c}{m_{e}^{2}c^{4}}
\frac{(k_{B}T)^{8}}{(\hbar c)^{6}}
\gamma_{p}^{6}e^{-\gamma_{p}}(1+\gamma_{p})
\langle {\rm block} \rangle_{\nu_{e}\bar{\nu}_{e}}^{\rm plas} ,
\eeq
where $\alpha_{\rm fine}\approx 1/137$ is the fine-structure constant and
$\gamma_{p} \approx 2\sqrt{(\alpha_{\rm fine}/9\pi) (\pi^{2}+3\eta_{e})}$.
The blocking factor is approximately given by
\beq
\langle {\rm block} \rangle_{\nu_{e}\bar{\nu}_{e}}^{\rm plas} \approx
\left[
1+\exp \left(
\eta_{\nu e} - 
\frac{\langle \epsilon_{\nu_{e}\bar{\nu}_{e}} \rangle^{\rm plas}}
{k_{B}T}
\right)
\right]^{-1}
\left[
1+\exp \left(
\eta_{\bar{\nu} e} - 
\frac{\langle \epsilon_{\nu_{e}\bar{\nu}_{e}} \rangle^{\rm plas}}
{k_{B}T}
\right)
\right]^{-1},
\eeq
where
\beq
\langle \epsilon_{\nu_{e}\bar{\nu}_{e}} \rangle^{\rm plas} =
\frac{k_{B}T}{2}\left(
2 + \frac{\gamma_{p}^{2}}{1+1\gamma_{p}}
\right)
\eeq
is the average energy of neutrinos produced by the plasmon decay.
The associated neutrino energy emission rate is given by
\beq
Q_{\nu_{e}\bar{\nu}_{e}}^{\rm plas} =
\frac{\rho}{m_{u}}\gamma_{\nu_{e}\bar{\nu}_{e}}^{\rm plas}
\langle \epsilon_{\nu_{e}\bar{\nu}_{e}} \rangle^{\rm plas}.
\eeq

Similarly, the number emission rate of $\nu_{x}$ or $\bar{\nu}_{x}$ by the
plasmon decay and the associated energy emission rate are given by
\beqn
&&
\gamma_{\nu_{x}\bar{\nu}_{x}}^{\rm plas} = 
\frac{m_{u}}{\rho}
\frac{(C_{V}-1)^{2}}{192 \pi^{3} \alpha_{\rm fine}}
\frac{\sigma_{0}c}{m_{e}^{2}c^{4}}
\frac{(k_{B}T)^{8}}{(\hbar c)^{6}}
\gamma_{p}^{6}e^{-\gamma_{p}}(1+\gamma_{p})
\langle {\rm block} \rangle_{\nu_{x}\bar{\nu}_{x}}^{\rm plas} , \\
&&
Q_{\nu_{x}\bar{\nu}_{x}}^{\rm plas} =
\frac{\rho}{m_{u}}\gamma_{\nu_{x}\bar{\nu}_{x}}^{\rm plas}
\langle \epsilon_{\nu_{x}\bar{\nu}_{x}} \rangle^{\rm plas}, 
\eeqn
where the average neutrino energy is 
$\langle \epsilon_{\nu_{x}\bar{\nu}_{x}} \rangle^{\rm plas} =
 \langle \epsilon_{\nu_{e}\bar{\nu}_{e}} \rangle^{\rm plas}$
and the blocking factor is given by
\beq
\langle {\rm block} \rangle_{\nu e}^{\rm pair} \approx
\left[
1+\exp \left(
\eta_{\nu x} - 
\frac{\langle \epsilon_{\nu_{x}\bar{\nu}_{x}} \rangle^{\rm pair}}
{k_{B}T}
\right)
\right]^{-1} 
\left[
1+\exp \left(
\eta_{\bar{\nu} x} - 
\frac{\langle \epsilon_{\nu_{x}\bar{\nu}_{x}} \rangle^{\rm pair}}
{k_{B}T}
\right)
\right]^{-1}.
\eeq

\subsection{Nucleon-nucleon bremsstrahlung}

We follow Burrows et al. \cite{BRT06} for the rate 
of pair creation of neutrinos by the nucleon-nucleon bremsstrahlung
radiation. They derived the neutrino energy emission
rate associated with the pair creation of $\nu_{x}$ or $\bar{\nu}_{x}$
by the nucleon-nucleon bremsstrahlung radiation without the blocking
factor:
\beq
Q_{\nu_{x}\bar{\nu}_{x}}^{\rm Brems, 0} =
3.62 \times 10^{5} \zeta^{\rm Brems} 
\left( X_{n}^{2} + X_{p}^{2} + \frac{28}{3}X_{n}X_{p} \right)
\rho^{2}
\left( \frac{k_{B}T}{m_{e}c^{2}} \right)^{4.5} 
\langle \epsilon_{\nu_{x}\bar{\nu}_{x}} \rangle^{\rm Brems}, 
\eeq 
where $\zeta^{\rm Brems} \sim 0.5$ is a correction factor and the
average energy is 
\beq
\langle \epsilon_{\nu_{x}\bar{\nu}_{x}} \rangle^{\rm Brems} 
\approx 4.36k_{B}T.
\eeq
We multiply $Q_{\nu_{x}\bar{\nu}_{x}}^{\rm Brems, 0}$ by the blocking
factor,
\beq
\langle {\rm block} \rangle_{\nu_{x}\bar{\nu}_{x}}^{\rm Brems} \approx
\left[
1+\exp \left(
\eta_{\nu x} - 
\frac{\langle \epsilon_{\nu_{x}\bar{\nu}_{x}} \rangle^{\rm Brems}}
{k_{B}T}
\right)
\right]^{-1}
\left[
1+\exp \left(
\eta_{\bar{\nu} x} - 
\frac{\langle \epsilon_{\nu_{x}\bar{\nu}_{x}} \rangle^{\rm Brems}}
{k_{B}T}
\right)
\right]^{-1},
\eeq
to obtain the 'blocked' neutrino energy emission rate
\beq
Q_{\nu_{x}\bar{\nu}_{x}}^{\rm Brems} = 
Q_{\nu_{x}\bar{\nu}_{x}}^{\rm Brems, 0}
\langle {\rm block} \rangle_{\nu_{x}\bar{\nu}_{x}}^{\rm Brems}.
\eeq
The number emission rate of $\nu_{x}$ or $\bar{\nu}_{x}$ is readily
given by
\beq
\gamma_{\nu_{x}\bar{\nu}_{x}}^{\rm Brems} =
3.62 \times 10^{5} \zeta^{\rm Brems} 
\left( X_{n}^{2} + X_{p}^{2} + \frac{28}{3}X_{n}X_{p} \right)
m_{u}\rho
\left( \frac{k_{B}T}{m_{e}c^{2}} \right)^{4.5} 
\langle {\rm block} \rangle_{\nu_{x}\bar{\nu}_{x}}^{\rm Brems}. 
\eeq 

Noting that the weak interaction coefficients of the
bremsstrahlung radiation are\cite{Itoh1996}
$(1-C_{V})^{2} + (1-C_{A})^{2}$ for the pair
creation of $\nu_{x}\bar{\nu}_{x}$ and 
$C_{V}^{2} + C_{A}^{2}$ for the pair
creation of $\nu_{e}\bar{\nu}_{e}$,
the number emission rate and the associated energy emission rate
for $\nu_{e}$ or $\bar{\nu}_{e}$ may be written as
\beqn
&&
\gamma_{\nu_{e}\bar{\nu}_{e}}^{\rm Brems} = 
\frac{C_{V}^{\ 2} + C_{A}^{\ 2}}
     {(1-C_{V})^{2} +(1-C_{A})^{2}}
 \gamma_{\nu_{x}\bar{\nu}_{x}}^{\rm Brems}, \\
&&
Q_{\nu_{e}\bar{\nu}_{e}}^{\rm Brems} = 
\frac{C_{V}^{\ 2} + C_{A}^{\ 2}}
     {(1-C_{V})^{2} + (1-C_{A})^{2}}
 Q_{\nu_{x}\bar{\nu}_x}^{\rm Brems}. 
\eeqn

\section{Neutrino diffusion rates} \label{A_nudiff}

We follow Ref.~\citen{RL03} for the diffusive neutrino-number emission
rate $\gamma_{\nu}^{\rm diff}$ and the associated energy emission rate
$Q_{\nu}^{\rm diff}$ appeared in Eqs. (\ref{Q_leak}) and (\ref{g_leak}). 
and present the explicit forms of them  in the following for 
convenience. An alternative definition of the diffusion rates 
are found in Ref.~\citen{RJS96}.

\subsection{Neutrino diffusion rates}
To calculate the neutrino diffusion rates 
$\gamma_{\nu}^{\rm diff}$ and $Q_{\nu}^{\rm diff}$, we first
define neutrino diffusion time.
In this paper, we consider cross sections 
for scattering on nuclei ($\sigma_{\nu A}^{\rm sc}$), 
on free protons ($\sigma_{\nu p}^{\rm sc}$),
and on free neutrons ($\sigma_{\nu n}^{\rm sc}$), and
for absorption on free nucleons $\sigma_{\nu n}^{\rm ab}$ 
for electron neutrinos and $\sigma_{\nu p}^{\rm ab}$ for electron
anti-neutrinos.

Ignoring the higher order correction terms in neutrino energy $E_{\nu}$, 
these neutrino cross sections can be written in general as
\beq
\sigma(E_{\nu}) = E_{\nu}^{2}\tilde{\sigma} , \label{crosssect}
\eeq
where $\tilde{\sigma}$ is a 'cross section' in which $E_{\nu}^{2}$
dependence is factored out.
In practice, the cross sections contain the correction terms which cannot be expressed
in the form of Eq. (\ref{crosssect}). We take account of these correction terms,
approximating neutrino-energy dependence by temperature dependence according to
\beq
E_{\nu} \approx k_{B}T\frac{F_{3}(\eta_{\nu})}{F_{2}(\eta_{\nu})}.
\eeq

The opacity is written as 
\beq
\kappa(E_{\nu}) = \sum \kappa_{i}(E_{\nu}) 
= E_{\nu}^{2} \sum \tilde{\kappa}_{i} = E_{\nu}^{2} \tilde{\kappa},
\eeq
and optical depth is calculated by
\beq
\tau (E_{\nu}) = \int \kappa (E_{\nu}) ds 
= E_{\nu}^{2}\int \tilde{\kappa} ds 
= E_{\nu}^{2} \tilde{\tau}.
\eeq
Then, we define neutrino diffusion time by
\beq
T_{\nu}^{\rm diff} (E_{\nu}) \equiv
a^{\rm diff} \frac{\Delta x(E_{\nu})}{c}\tau(E_{\nu})
=
E_{\nu}^{2} a^{\rm diff} \frac{\tilde{\tau}^{2}}{c\tilde{\kappa}} 
= E_{\nu}^{2} \tilde{T}_{\nu}^{\rm diff},
\eeq
where the distance parameter $\Delta x (E_{\nu})$ is given by
\beq
\Delta x (E_{\nu}) = \frac{\tau (E_{\nu})}{\kappa (E_{\nu})}. 
\eeq
Note that $\tilde{T}_{\nu}^{\rm diff}$ can be calculated only using
matter quantities.
$a^{\rm diff}$ is a parameter which controls how much neutrinos
diffuse outward and we chose it to be $3$ following Ref.~\citen{RJS96}.
For a larger value of $a^{\rm diff}$, corresponding neutrino emission
rate due to diffusion becomes smaller.

Finally, we define the neutrino diffusion rates by
\beqn
&& \gamma_{\nu}^{\rm diff} \equiv \frac{m_u}{\rho}
\int \frac{n_{\nu}(E_{\nu})}{T_{\nu}^{\rm diff}(E_{\nu})}dE_{\nu}
= \frac{1}{a^{\rm diff}} \frac{m_u}{\rho} 
  \frac{4\pi c g_{\nu}}{(hc)^{3}}
  \frac{\tilde{\kappa}}{\tilde{\tau}^{2}} T F_{0}(\eta_{\nu}), \\
&& Q_{\nu}^{\rm diff} \equiv 
\int \frac{E_{\nu}n_{\nu}(E_{\nu})}{T_{\nu}^{\rm diff}(E_{\nu})}dE_{\nu}
= \frac{1}{a^{\rm diff}}\frac{4\pi c g_{\nu}}{(hc)^{3}}
  \frac{\tilde{\kappa}}{\tilde{\tau}^{2}} T^{2} F_{1}(\eta_{\nu}).
\eeqn

%
\subsection{Summary of cross sections} \label{CS}

In this subsection, we briefly summarize the cross sections adopted in the
present neutrino leakage scheme.

\subsubsection{Neutrino nucleon scattering}

The total $\nu$-$p$ scattering cross section $\sigma_{p}$ for all
neutrino species is given by
\beq
\sigma_{\nu p}^{\rm sc} =
\frac{\sigma_{0}}{4}\left(\frac{E_{\nu}}{m_{e}c^{2}}\right)^{2} 
\left[(C_{V}-1)^{2} + 3g_{A}^{2} (C_{A}-1)^{2} \right],
\eeq
where $g_{A}$ is
the axial-vector coupling constant $g_{A} \approx -1.26$.
On the other hand, the total $\nu -n$ scattering cross section
$\sigma_{n}$ is
\beq
\sigma_{\nu n}^{\rm sc} =
\frac{\sigma_{0}}{16}\left(\frac{E_{\nu}}{m_{e}^{2}c^{2}}\right)^{2} 
\left[ 1 + 3g_{A}^{2} \right] .
\eeq

\subsubsection{Coherent scattering of neutrinos on nuclei}

\begin{figure}[t]
  \begin{center}
    \includegraphics[scale=1.0]{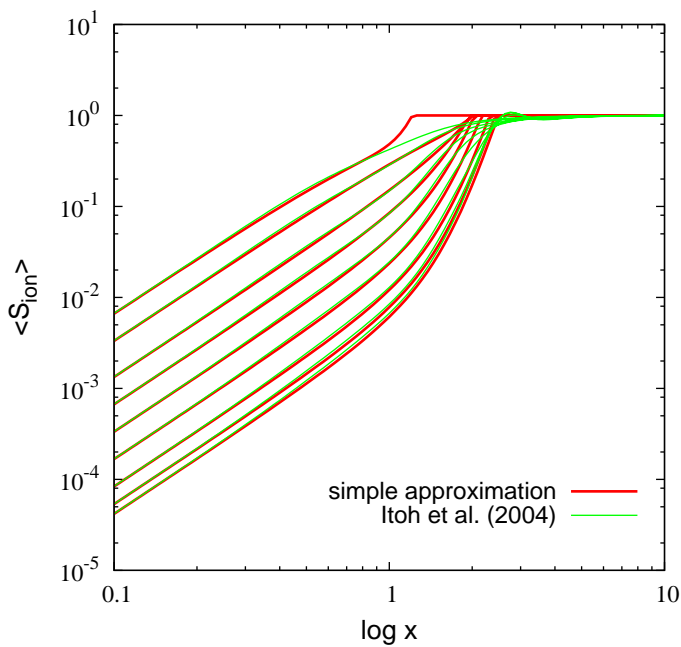} \\
  \end{center}
  \caption{Comparison of the ion-ion correction term, as a function of 
  $x \equiv E_{\nu}a_{I}/(\hbar c)$, between our simplified 
  estimation and a detailed fitting formula by Itoh et al. From the top
 to the bottom, the curves are for $\Gamma=1$, 2, 5, 10, 20, 40, 80,
 125, 160.}\label{fig_ion}
\end{figure}

The differential cross section for the $\nu$-$A$ neutral current
scattering is written as\cite{CrossS}
\beq
\frac{d\sigma_{A}^{\rm sc}}{d\Omega} = \frac{\sigma_{0}}{64 \pi}
\left(\frac{E_{\nu}}{m_{e}c^{2}}\right)^{2}
A^{2}\left[{\cal W}{\cal C}_{\rm FF} + {\cal C}_{\rm LOS}\right]^{2}
\langle {\cal S}_{\rm ion} \rangle
(1+ \cos\theta),
\eeq
where $\theta$ is the azimuthal angle of the scattering and
\beq
{\cal W} = 1- \frac{2Z}{A}(1-2\sin^{2}\theta_{W}).
\eeq
$\langle {\cal S}_{\rm ion} \rangle$, ${\cal C}_{\rm LOS}$, and 
${\cal C}_{FF}$ are correction factors due to 
the Coulomb interaction between the nuclei,\cite{Horowitz97}
due to the electron polarization,\cite{Leinson88}
and due to the finite size of heavy nuclei\cite{FsizeN}.
Because it is known that the correction factor ${\cal C}_{\rm LOS}$ is
important only for low neutrino energies\cite{BRT06}, 
we consider only $\langle {\cal S}_{\rm ion} \rangle$ and $C_{FF}$.

The correction factor due to the Coulomb interaction between the nuclei
is given by
\beq
\langle {\cal S}_{\rm ion} \rangle = 
\frac{3}{4}\int_{-1}^{1}d\cos \theta (1+\cos\theta) (1- \cos \theta) S_{\rm ion}.
\eeq
Itoh et al.\cite{Itoh2004} presented a detailed fitting formula for the
correction factor. However, the fitting formula is very complicated, we
use a simple approximation based on Ref.~\citen{Itoh75}.

$S_{\rm ion}$ can be written approximately as\cite{Itoh75}
\beq
S_{\rm ion} \approx \frac{(q a_{I})^{2}}{3\Gamma + f(\Gamma)(q a_{I})^{2}}
\eeq
for $(q a_{I} \ll 1)$, where $q = (2E_{\nu}/\hbar c) \sin (\theta /2)$,
$a_{I} = (4\pi n_{A}/3)^{-1/3}$ is the ion-sphere radius, $n_{A}$ is
the number density of a nucleus, and 
$\Gamma = (Ze)^{2}/(a_{I}k_{B}T)$ is the conventional parameter that 
characterizes the strongness of the Coulomb interaction.
$f(\Gamma)$ is given by\cite{Itoh2004}
\beq
f(\Gamma) \approx 0.73317 - 0.39890 \Gamma + 0.34141 \Gamma^{1/4}
+ 0.05484 \Gamma^{-1/4}.
\eeq
The integration approximately gives for 
$x \equiv E_{\nu}a_{I}/(\hbar c) < 1$
\beq
\langle {\cal S}_{\rm ion} \rangle \approx
 \frac{1}{6  }\frac{1        }{\Gamma    }x^{2}
-\frac{1}{30 }\frac{f(\Gamma)}{\Gamma^{2}}x^{4}
+\frac{1}{135}\frac{(f(\Gamma))^{2}}{\Gamma^{3}}x^{6}
-\frac{1}{567}\frac{(f(\Gamma))^{3}}{\Gamma^{4}}x^{8}
+\frac{1}{2268}\frac{(f(\Gamma))^{4}}{\Gamma^{5}}x^{10}.
\eeq
To use this expression for the case of 
$x \ge 1$, 
we set the maximum value as $ \langle {\cal S}_{\rm ion} \rangle = 
{\rm max} (1, \langle {\cal S}_{\rm ion} \rangle)$ where 
$\langle {\cal S}_{\rm ion} \rangle =1 $ corresponds to the
case without the correction.
Note that $x$ can be larger than 
unity\footnote{$x \sim (E_{\nu}/27 {\rm MeV})(\rho_{12}/A)^{-1/3}$ where
$\rho_{12}$ is the rest-mass density in units of $10^{12}$ g/cm$^{3}$}.
In this case, the simple approximation on average underestimates the 
effect of the Coulomb interaction between the nuclei (see Fig. \ref{fig_ion}).

Figure \ref{fig_ion} compares the correction term as a function of the
parameter $x$ calculated by our simple approximation and by 
a detailed fitting formula by Itoh et al.\cite{Itoh2004}
For smaller values of $\Gamma$, the disagreements become prominent.
Note that the typical value of $\Gamma$ inside the collapsing core with $T\sim 1$
MeV, $\rho \sim 10^{12}$ g/cm$^{3}$,  $A=56$ and $Z=26$ ($^{56}$Fe) is
$\Gamma \sim 47$.

\subsubsection{Absorption on free neutrons}
The total cross section of the absorption of electron neutrinos on free neutrons
is given by\cite{CrossS}
\beq
\sigma_{n}^{\rm ab} = \sigma_{0} \left(\frac{1+3g_{A}^{2}}{4}\right)
\left(\frac{E_{\nu} + \Delta_{\rm np}}{m_{e}c^{2}}\right)^{2}
\left[1- \left(\frac{m_{e}c^{2}}{E_{\nu}+\Delta_{\rm
np}}\right)\right] W_{M},
\eeq
where $\Delta_{\rm np} = m_{n}c^{2}-m_{p}c^{2}$, and
$W_{M}$ is the correction for weak magnetism and recoil which is approximately
given by 
\beq
W_{M} \approx 1 + 1.1 \frac{E_{\nu}}{m_{n}c^{2}}.
\eeq

Similarly, the total cross section of the absorption of electron anti-neutrinos on 
free protons is given by\cite{CrossS}
\beq
\sigma_{p}^{\rm ab} = \sigma_{0} \left(\frac{1+3g_{A}^{2}}{4}\right)
\left(\frac{E_{\bar{\nu}} - \Delta_{\rm np}}{m_{e}c^{2}}\right)^{2}
\left[1- \left(\frac{m_{e}c^{2}}{E_{\bar{\nu}}-\Delta_{\rm
np}}\right)\right] W_{\bar{M}},
\eeq
where $W_{\bar{M}}$ is approximately given by 
\beq
W_{\bar{M}} \approx 1 - 7.1 \frac{E_{\bar{\nu}}}{m_{p}c^{2}}.
\eeq

%

\end{document}